\pdfoutput=1
\RequirePackage{ifpdf}
\ifpdf 
\documentclass[pdftex]{sigma}
\else
\documentclass{sigma}
\fi

\numberwithin{equation}{section}

\begin{document}

\allowdisplaybreaks

\renewcommand{\thefootnote}{$\star$}

\renewcommand{\PaperNumber}{089}

\FirstPageHeading

\ShortArticleName{Nonregular Separation of Variables}

\ArticleName{Solutions of Helmholtz and Schr\"odinger Equations\\ with  Side Condition and Nonregular Separation\\ of Variables\footnote{This
paper is a contribution to the Special Issue ``Symmetries of Dif\/ferential Equations: Frames, Invariants and Applications''. The full collection is available at \href{http://www.emis.de/journals/SIGMA/SDE2012.html}{http://www.emis.de/journals/SIGMA/SDE2012.html}}}

\Author{Philip BROADBRIDGE~$^\dag$, Claudia M.~CHANU~$^\ddag$ and Willard MILLER~Jr.~$^\S$}

\AuthorNameForHeading{P.~Broadbridge, C.M.~Chanu and  W.~Miller~Jr.}

\Address{$^\dag$~School of Engineering and Mathematical Sciences,
 La Trobe University, Melbourne, Australia}
\EmailD{\href{mailto:P.Broadbridge@latrobe.edu.au}{P.Broadbridge@latrobe.edu.au}}

\Address{$^\ddag$~Dipartimento di Matematica G.~Peano,  Universit\`a di Torino, Torino, Italy}
\EmailD{\href{mailto:claudiamaria.chanu@unito.it}{claudiamaria.chanu@unito.it}}

\Address{$^\S$~School of Mathematics,  University of Minnesota,
 Minneapolis, Minnesota, 55455, USA}
\EmailD{\href{mailto:miller@ima.umn.edu}{miller@ima.umn.edu}}
\URLaddressD{\url{http://www.ima.umn.edu/~miller/}}

\ArticleDates{Received September 21, 2012, in f\/inal form November 19, 2012; Published online November 26, 2012}

\Abstract{Olver and Rosenau studied  group-invariant solutions of (generally nonlinear) partial dif\/ferential equations through the
 imposition of a side condition.
We apply a similar idea to the special case of f\/inite-dimensional Hamiltonian systems, namely Hamilton--Jacobi, Helmholtz and time-independent Schr\"odinger equations
with potential on $N$-dimensional Riemannian and pseudo-Riemannian manifolds, but with a linear side condition, where more structure is available. We show that the requirement of $N-1$ commuting
second-order symmetry operators, modulo a second-order linear  side condition corresponds  to nonregular separation of variables in an orthogonal coordinate system,
 characterized by a generalized St\"ackel matrix. The coordinates and solutions obtainable through true nonregular separation are distinct from those arising
through regular separation
 of variables. We develop the theory for these systems and provide  examples.}


\Classification{35Q40; 35J05}

\renewcommand{\thefootnote}{\arabic{footnote}}
\setcounter{footnote}{0}

\section{Introduction}

The primary motivation for this paper was the construction by Olver and Rosenau of group-invariant solutions of (generally nonlinear) partial dif\/ferential equations through the imposition of a
side condition~\cite{OR87, OR86}. Various forms of conditional symmetry methods, including  nonclassical  symmetry reduction, have been applied by several authors, e.g.~\cite{AHB,BC2,BC1,CM, LW,NC,Ovs}.
 We are interested in using linear second-order side conditions and separation of variables to f\/ind explicit solutions of Hamilton--Jacobi, Helmholtz, Laplace, wave and heat equations.
 We  exploit the special properties of f\/inite
order classical Hamiltonian systems and their quantum analogues represented by Schr\"odinger equations  to obtain new results  on separation of variables. The present paper is devoted to Hamilton--Jacobi and Helmholtz or time independent
Schr\"odinger equations with potential but the ideas are clearly applicable to more general Hamiltonian systems and to non-Hamiltonian systems such as dif\/fusion equations that share some features with time-dependent Schr\"odinger equations.

An example is the Schr\"odinger equation $H\Psi=E\Psi$ where $H=\Delta+V({\bf x})$ and $\Delta$ is the Laplace--Beltrami operator on some Riemannian or
pseudo-Riemannian manifold.
We  look for solutions of this equation that also satisfy a side condition $S\Psi=0$ where $S$ is some given linear partial dif\/ferential operator in the
 variables~$\bf x$. A consistency condition for the existence of nontrivial solutions $\Psi$ is  $[H,S]=AS$ for some linear partial dif\/ferential operator~$ A$.
 Moreover a~linear dif\/ferential operator~$L$ will be a  symmetry operator for $H$, (modulo $S\Psi=0$) if $[H,L]=BS$ for some linear partial dif\/ferential operator~$B$.
 To make contact with solutions that are separable in some system of coordinates,
 we restrict to the case where the symmetries and the side condition are second-order partial dif\/ferential operators.

  A Hamilton--Jacobi analog is the equation ${\cal H}=E$ where, in orthogonal coordinates,
\[
  {\cal H}=\sum_{j=1}^Ng^{jj}({\bf x})p_j^2+V({\bf x}),
\]
  which is double the classical Hamiltonian.
We look for solutions $u({\bf x})$ of the Hamilton--Jacobi equation, where $p_j=\partial_{x^j}u$, subject to the side condition ${\cal S}({\bf x},{\bf p})=0$.
The consistency requirement is the Poisson bracket relation $\{{\cal H},{\cal S}\}={\cal A}{\cal S}$ for some function ${\cal A}$ on phase space. Moreover a phase space
function~$\cal L$ will be a  constant of the motion for $\cal H$, (modulo ${\cal S}=0$) if $\{{\cal H},{\cal L}\}={\cal B}{\cal S}$ for some phase space function~$\cal B$.

Finding such systems directly from their def\/inition leads to great computational complexity. We explore a new method, based on a generalization of St\"ackel form, i.e., a generalization of separable systems corresponding tho a St\"ackel matrix,
that allows us to generate such systems ef\/f\/iciently.

The second motivation for this paper is the general theory of separation of variables for both linear and nonlinear
 partial dif\/ferential equations~\cite{EGKWM83,KMILW93,MIL83+,MIL83,EGKWM83+}. In these works the authors point out that there are two types of variable separation: regular and nonregular.
 Regular separation is the most familiar and was exploited by pioneers such as St\"ackel~\cite{Stackel1891} and Eisenhart~\cite{EIS48, EIS34}, \cite[Appendix~13]{EIS49}.
 For regular orthogonal separation of a Helmholtz or Schr\"odinger equation on an $N$-dimensional manifold there are always $N$ separation
constants and the associated separable solutions form a basis for the solution space. There is a well developed theory for regular orthogonal separation of these equations,
 including  classif\/ication of possible separable coordinate systems in various constant curvature spaces and intrinsic characterizations of the separable systems,
see for example~\cite{BF1980,ERNIE, KM1983,KMJ, KMJ87,LC,SHAP+, SHAP} in addition to earlier cited references.

For nonregular separation, on the other hand, the number of separation constants is strictly less than $N$ and the separable solutions do not form a basis.
 Symmetry adapted solutions of partial dif\/ferential equations
are prominent examples of this class, but except for these special solutions there is virtually no structure or classif\/ication theory.
First attempts of geometric interpretation of nonregular separation are given in \cite{HJ0, IMA06,CR, DegRast} where nonregular separation is considered as separation
in which the separated solution must satisfy additional constraints  that can be seen as side conditions for the equation.
In this paper, however,
we will show that solutions of Helmholtz and Schr\"odinger equations with
  second-order
side conditions provide a class of nonregular orthogonal  separation of variables
that can be characterized intrinsically. Further, each of these systems is associated with a generalized St\"ackel matrix and this association enables us to generate
nonregular separable systems very easily.

In Sections~\ref{HJreview} and~\ref{Helmholtzreview} we review the St\"ackel construction for regular additive separation of Hamilton--Jacobi equations and
regular multiplicative separation and $R$-separation for Helmholtz equations, and
 make some comments on
their geometric characterizations. We also discuss the ef\/fect of adding vector and scalar
potentials. In Section~\ref{genstac} we introduce a  generalized St\"ackel matrix with one arbitrary column and show that its use leads to additive separable solutions
 of the Hamilton--Jacobi equation, with a side condition. We express the results in a Hamiltonian formalism. Then in Section~\ref{Helmholtzstackel} we carry out the
analogous construction for Helmholtz and Schr\"odinger eigenvalue equations. Section~\ref{nonregsep2} is the main theoretical contribution of our paper. We show that the requirement of maximal
nonregular separation for Hamilton--Jacobi equations is equivalent to separation with a  generalized St\"ackel matrix and can be characterized geometrically. With some modif\/ications,
 the same is true of $R$-separation for Helmholtz and Schr\"odinger equations (though there is still  a ``generalized Rodrigues form'' gap in the geometrical characterization).  Section~\ref{examples} is devoted to examples of nonregular separation and discussion of their various types and
signif\/icance. We prove a ``no go'' theorem to the ef\/fect that nonregular $R$-separation does not occur for the Helmholtz (or Schr\"odinger) equation with no
potential or scalar potential on a 2D Riemannian or pseudo-Riemannian manifold. However 2D nonregular separation can occur for equations with vector or magnetic
potentials. Section~\ref{vector potential} develops the theory for
 two-dimensional
systems with vector potential. The self adjoint Schr\"odinger equation for a charged particle
in two spatial dimensions, interacting with a classical  electromagnetic f\/ield, again has no new separable  coordinate systems that are obtainable from
a generalised St\"ackel matrix. However, nonself-adjoint equations such as the analogous solute transport equation with f\/irst-order convective terms replacing
magnetic potential terms, do indeed have new nonregular separable coordinate systems.

 We provide examples of nonregular $R$-separation for various zero potential, scalar potential and vector potential 3D systems: Euclidean space, Minkowski space and
nonzero constant curvature space, including a Euclidean space example due to Sym~\cite{SYM} of nonregular separation with two side conditions. The f\/inal Section~\ref{conclusions} sums up our conclusions and points the way for future research on nonregular separation. The coordinate systems and solutions for true nonregular separation, i.e., nonregular separation for which regular separation doesn't occur,
are distinct from those for regular separation and further study of their scope and signif\/icance is in order.

\section[Review of regular orthogonal separation for the Hamilton-Jacobi equation ${\cal H}=E$]{Review of regular orthogonal separation\\ for the Hamilton--Jacobi equation $\boldsymbol{{\cal H}=E}$}\label{HJreview}

Write  the Hamilton--Jacobi equation  in terms of $u_i=\partial_iu(\bf x)$, as
\begin{gather}\label{HJ}
 {\cal H}\equiv\sum_{i=1}^N H^{-2}_i u^2_i+V({\bf x})=E.
\end{gather}
Here,  the metric in the orthogonal coordinates $x^i$ is
$
ds^2=\sum\limits_{i=1}^NH_i^2(dx^i)^2.
$
We want to obtain additive separation, so that $ \partial_j
u_i\equiv \partial_j\partial_iu = 0$ for $i\ne j$.
 Requiring that the solution $u$ depends on $n$ parameters $(\lambda_1,\ldots, \lambda_N)$ implies the existence
of separation equations in the form
\begin{gather}\label{sepeqns1} u_i^2        + v_i\big(x^i\big)+\sum_{j=1}^N s_{ij}\big(x^i\big)\lambda_j=0,\qquad
i=1,\dots,N,\qquad \lambda_1=-E.
\end{gather}
Here $\partial_k s_{ij}(x^i)=0$ for $k\ne i$ and $\det (s_{ij})\ne 0 $.
We say that $S=(s_{ij})$ is a {\it St\"ackel matrix}. Set $T=S^{-1}$.

Then (\ref{HJ}) can be recovered from (\ref{sepeqns1}) provided
$H_j^{-2}=T^{1j}$ and $V=\sum_jv_jT^{1j}$.
The quadratic forms $ {\cal L}^{\ell}=\sum\limits_{j=1}^N T^{\ell j} (u_j^2+v_j)$ satisfy
${\cal L}^{\ell}=-\lambda_{\ell} $
 for a separable solution.
Furthermore, setting $u_i = p_i$, we claim
\begin{gather}\label{commident1}
\big\{ {\cal L}^{\ell}, {\cal L}^j\big\}=0,\qquad \ell \ne j \end{gather}
where
$\{ {\cal H},{\cal K}\}=\sum\limits_{i=1}^N (\partial_{x^i} {\cal K} \partial_{ p_i} {\cal H} -\partial_{ x^i} {\cal H} \partial_{ p_i} {\cal K})
$
is the Poisson Bracket. Thus the ${\cal L^{\ell} }$, $2\le \ell\le N$, are
{\it constants of
the motion} for the {\it Hamiltonian } ${\cal H}= {\cal L}^{(1)} $.  For the proof of (\ref{commident1})
one notes that
\[
 \sum_{j=1}^NT^{\ell j}s_{jk}\big(x^j\big)=\delta_{\ell k},
 \]
Dif\/ferentiating this identity with respect to $x^i$, we f\/ind
\[
 \sum_{j=1}^N\partial_{i}T^{\ell j}  s_{jk}\big(x^j\big)+T^{\ell i} s'_{ik}\big(x^i\big)=0,
 \]
so
\begin{gather*}
\partial_{i}T^{\ell j}=-T^{\ell i}\sum_{k=1}^N s'_{ik}T^{k j}.
\end{gather*}
We substitute this expression  into the left hand side of~(\ref{commident1}) and obtain the desired result after a~routine computation.

\section{Review of the St\"ackel procedure for the Helmholtz\\ or Schr\"odinger equation}\label{Helmholtzreview}

We can perform an analogous construction of eigenfunctions  for a Helmholtz  operator, using the  St\"ackel matrix~$S$. We demand eigenfunctions of~$H$ in the separated form $\Psi=\prod\limits_{j=1}^N\Psi^{(j)}(x^j)$
 and depending on the maximal number of parameters. Then, the separation equations are of the form
\begin{gather}\label{prodsep0} \partial_\ell^2\Psi+f_\ell(x^\ell)\partial_\ell\Psi+\left(v_\ell\big(x^\ell\big)-\sum_{j=1}^{N}s_{\ell j}\big(x^\ell\big)\lambda_j\right)\Psi=0,\qquad \ell=1,\dots,N
\end{gather}
 for suitable functions $f_\ell$, $v_\ell$ to be determined.
Thus we have the eigenvalue equations
\begin{gather*}
L_k\Psi\equiv \sum_{\ell=1}^NT^{k\ell}\big(\partial_\ell^2+f_\ell\partial_\ell+v_\ell\big)\Psi=\lambda_k\Psi,\qquad k=1,\dots,N,
\end{gather*}
where $L_1=H$.
Based on our calculations of the preceding section, we  can establish the commutation relations
\begin{gather*}
[L_s,L_t]=0.
\end{gather*}

More generally we can consider $R$-separation for a general Helmholtz equation. In local coordinates $z^j$ on an $N$-dimensional pseudo-Riemannian manifold this equation takes the invariant form
\begin{gather}\label{generalHelmholtz}
H\Theta\equiv \left(\Delta_N+\sum_{j=1}^N F^{j}\partial_j+V\right)\Theta=E\Theta,
\end{gather}
where
\[
\Delta_N\equiv \frac{1}{\sqrt{g}}\sum_{j,k=1}^N \partial_j\big(g^{jk}\sqrt{g}\partial_k\big)
\]
is the Laplace--Beltrami operator. We say that this equation is {\it  $R$-separable} in local orthogonal coordinates $x^j$ if  there is a f\/ixed nonzero function $R({\bf x})$ such that (\ref{generalHelmholtz}) admits solutions
\[
\Theta=\exp(R)\Psi=\exp(R)\prod_{j=1}^N\Psi^{(j)}\big(x^j\big),
\]
where  $\Psi$
 is a regular separated solution, i.e., it
satisf\/ies the separation equations~(\ref{prodsep0}).
In this case the symmetry operators are ${\tilde L}_k=\exp(R)L_k\exp(-R)$ and equations (\ref{prodsep0}) become
\begin{gather*}
\partial_\ell^2\Theta+\big(f_\ell\big(x^\ell\big)-2\partial_\ell R\big)\partial_\ell\Theta+\left(v_\ell\big(x^\ell\big)-\partial_{\ell\ell}R+(\partial_\ell R)^2-\sum_{j=1}^{N}s_{\ell j}\big(x^\ell\big)\lambda_j\right)\Theta=0,
\end{gather*}
$ \ell=1,\dots,N$. Then we have
\begin{gather*}
 [{\tilde L}_s,{\tilde L}_t]=0,
 \end{gather*}
where ${\tilde L}_1=H$.

Now consider the case
\begin{gather}\label{nomagnetic}
{\tilde H}\Theta\equiv (\Delta_N+V)\Theta=E\Theta,
\end{gather}
 i.e., the case where there is no magnetic f\/ield, and $V$ is real.  We can def\/ine an inner product on the space of $C^\infty$ real valued functions $f^{(1)}({\bf z})$, $f^{(2)}({\bf z})$ with compact support in~${\mathbb R}^N$, with respect to which $\tilde H$ is formally self-adjoint
\[
\langle f^{(1)},f^{(2)}\rangle =\int_{{\mathbb R}^N}f^{(1)}({\bf z})f^{(2)}({\bf z})\sqrt{g({\bf z})}\ d{\bf z}.
\]
If $L=\sum\limits_{j,k=1}^Na^{jk}({\bf z})\partial^2_{jk} +\sum\limits_{\ell=1}^Nh_\ell({\bf z})\partial_\ell+W({\bf z})$ is a real symmetry operator then it can be uniquely decomposed as $L=L^{(1)}+L^{(2)}$ where  $L^{(1)}$ is formally self-adjoint and $L^{(2)}$ is formally skew-adjoint
\[
L^{(1)}= \frac{1}{\sqrt{g}}\sum_{j,k=1}^N \partial_j\big(a^{jk}\sqrt{g}\partial_k\big)+{\tilde W},\qquad L^{(2)}= \frac{1}{\sqrt{g}}\sum_{\ell=1}^N\left({\tilde h}_\ell\partial_\ell
+\frac12\partial_\ell {\tilde h}\right).
\]
Moreover, both $L^{(1)}$, $L^{(2)}$ are symmetry operators. Note that we have $L^{(2)}=0$ unless $\tilde H$ admits a f\/irst-order symmetry operator.

Suppose the system admits $N$ algebraically independent commuting symmetry opera\-tors~${\tilde L}_s$ such that the coef\/f\/icients $(a^{jk}_s)$ of the
second-order terms in the symmetries admit a basis of common eigenforms. (Without loss of generality we can restrict to the self-adjoint case
${\tilde L}_s={\tilde L}_s^{(1)}$.)   Then it is well established~\cite{KMJ87}, from the examination of the third-order terms in the relations
$[{\tilde L}_s,{\tilde L}_t]=0$ that there is an orthogonal coordinate system~$x^j$ and corresponding St\"ackel matrix $ \left(s_{jk}(x^j)\right)$ such that
\[
{\tilde H}=\sum_{j=1}^N \frac{1}{h}\partial_j\big(hH_j^{-2}\partial_j\big) +V,\qquad h=H_1H_2\cdots H_N.
\]
Since the metric is in St\"ackel form, it is
 straightforward to verify that the symmetries  can be rewritten as
\[
 {\tilde L}_s=\sum_{j=1}^N T^{sj}\left(\partial^2_j+\left(\partial_j\frac{h}{S}\right)\partial_j~{+v_j}\right) ,
 \]
making it  evident that the f\/irst derivative terms are a gradient. (Here, $S$ is the determinant of the St\"ackel matrix.) Thus via an $R$-transform we can express our
system in the form
\[
H=\sum_{j=1}^N H_j^{-2}\partial^2_j+ {\hat V}.
\]
Then the third derivative  terms in the commutation relations $[L_s,L_t]=0$ are unchanged and the  cancellation of second derivative terms is satisf\/ied identically.
The f\/irst derivative terms just tell us that the transformed potential $\hat V$ is a St\"ackel multiplier, so that it permits separation in the coordinates~$x^j$.
 The zero-th order relation is satisf\/ied identically. Thus the  integrable system~(\ref{nomagnetic}) is $R$-separable. Under the same assumptions,
 but with a magnetic term added, this is no longer necessarily true. It is easy to see that if there is a function $G$ such that $F^{j}=\partial_jG$, i.e.,
 if the magnetic potential is a gradient, then the system is again $R$-separable. However, if the magnetic potential is not a gradient then it is no longer
necessarily true that integrability implies $R$-separability, even though the second-order terms in the Laplacian admit a common basis of eigenforms.
 See \cite{BW2004} for some examples.

\section{A generalization of St\"ackel form}\label{genstac}

We def\/ine a $N\times N$ generalized St\"ackel matrix by
\begin{gather}\label{genstackel1} S=  \begin{pmatrix} s_{11}\big(x^1\big)& s_{12}\big(x^1\big)& \cdots&s_{1,N-1}\big(x^1\big)& a_1({\bf x})\\
s_{21}\big(x^2\big)& s_{22}\big(x^2\big)& \cdots&s_{2,N-1}\big(x^2\big)& a_2({\bf x})\\
\cdots&\cdots&\cdots&\cdots&\cdots\\
s_{N1}\big(x^N\big)& s_{N2}\big(x^N\big)& \cdots&s_{N,N-1}\big(x^N\big)& a_N({\bf x})\end{pmatrix}.
\end{gather}
where the $a_i$ are arbitrary analytic functions of the variables $x^1,\dots,x^N$. We require that $S$ is a~nonsingular matrix. Set $T=S^{-1}$.
Now we  assume existence of
separation equations in the form
\begin{gather}\label{sepeqns2}
u_i^2 + v_i\big(x^i\big)+\sum_{\xi=1}^{N-1} s_{i\xi}\big(x^i\big)\lambda_\xi=0,\qquad
i=1,\dots,N,\qquad \lambda_1=-E.
\end{gather}
(Here Latin indices take values $1,\dots,N$ and Greek indices take values $1,\dots,N-1$.) Note that the term with $\lambda_N$ is missing.
Thus the general separated solution~$u$ will depend on~$N$ parameters (rather than~$N+1$), an example of (maximal)  nonregular separation.
Note that  equations~(\ref{sepeqns2}) can be considered as the restriction to the case $\lambda_N=0$ of
\begin{gather*}
u_i^2 + v_i\big(x^i\big)+\sum_{\xi=1}^{N-1} s_{i\xi}\big(x^i\big)\lambda_\xi+\lambda_N a_i(\mathbf x)=0,
\end{gather*}
which are not separated for $\lambda_N=0$. However, any solution of them  is a solution of the $N$ equations
\begin{gather}\label{constmotion1}
{\cal L}^{\ell}\equiv\sum_{j=1}^N T^{\ell j} \big(u_j^2+v_j\big)=\lambda_\ell,\qquad 1\le \ell\le N,\qquad {\cal L}^N=0,
\end{gather}
where ${\cal H}={\cal L}^{1}$ is  the Hamiltonian.
Hence, by solving (\ref{sepeqns2}) we get a separated solution of ${\cal H}=E$  satisfying also ${\cal L}^N=0$ as a side condition.
This construction shows that separation with a~generalized St\"ackel matrix means (nonregular) separation with a side condition.
In this case, functions~(\ref{constmotion1}) become our ``restricted constants of the motion'' ${\cal L}^{\alpha}=\lambda_\alpha$ (modulo ${\cal L}^N=0$).

Since
\[
\sum_{j=1}^NT^{\ell j}s_{jk}=\delta_{\ell k},
\]
dif\/ferentiating this identity with respect to $x^i$, gives
\[
 \sum_{j=1}^N\partial_{ i}T^{\ell j}\ s_{j\xi}(x_j)+T^{\ell i}\ s'_{i\xi}\big(x^i\big)=0,
\]
and
\[
 \sum_{j=1}^N\big(\partial_{i}T^{\ell j}\ s_{jN}+T^{\ell j}\partial_{i} a_j\big)=0,
 \]
so
\begin{gather}\label{Sinverse2}
\partial_{i}T^{\ell j}+T^{\ell i}\sum_{\xi=1}^{N-1} s'_{i\xi}T^{\xi j}+T^{N j}\sum_{h=1}^NT^{\ell h} \partial_{i} a_h=0.
\end{gather}
Using this result it is a straightforward computation to verify the Poisson bracket relations:
\begin{gather}\label{pb1}
\big\{{\cal L}^i,{\cal L}^j\big\}
=\left(\sum_{k,h=1}^N\big(T^{i k}T^{j h}-T^{j k}T^{i h}\big)\frac{\partial a_k}{\partial x^h}p_h\right){\cal L}^N.
\end{gather}
Relations (\ref{pb1})
can be considered as the consistency conditions that guarantee  the ${\cal L}^i$ are constants of
the motion for the Hamiltonian ${\cal H}$, modulo the side condition ${\cal L}^N=0$.
We have verif\/ied the relations
\[
\big\{{\cal L}^i,{\cal L}^j\big\}_{{\cal L}^N=0}=0,\qquad i,j=1,\dots,N
\]
for ${\cal L}^1={\cal H}$ and linearly independent quadratic forms ${\cal L}^i$.  Thus our construction has shown that separation with a
generalized St\"ackel matrix implies the existence of $N$ independent constants of motion in involution (modulo the side condition) diagonalised in the separable coordinates.

We can generalize Eisenhart's treatment of St\"ackel form in which he represented the quadratic forms $T^{\ell i}$ in terms of their eigenvalues with respect to the
metric $T^{1j}=H^{-2}_j$: $T^{\ell j}=\rho^{(\ell)}_jH^{-2}_j$. Here, $\rho^{(1)}_j=1$.  Then~(\ref{Sinverse2}) can be rewritten as a system of
 partial dif\/ferential equations for the~$\rho^{(\ell)}_j$:
\begin{gather}\label{T1}
\partial_i \rho^{(\ell)}_j+\rho^{(\ell)}_j\frac{\partial_i H^{-2}_j}{H^{-2}_j}+\rho^{(\ell)}_i H^{-2}_i\sum_{\xi=1}^{N-1}s'_{i\xi}\rho^{(\xi)}_j+\rho^{(N)}_j\sum_{h=1}^N\rho^{(\ell)}_h H^{-2}_h\partial_ia_h=0.
\end{gather}
In the special case $\ell=1$ these equations reduce to
\begin{gather*}
\frac{\partial_i H^{-2}_j}{H^{-2}_j}+ H^{-2}_i\sum_{\xi=1}^{N-1}s'_{i\xi}\rho^{(\xi)}_j+\rho^{(N)}_j\sum_{h=1}^N H^{-2}_h\partial_ia_h=0.
\end{gather*}
Substituting this result in (\ref{T1}) we obtain
\begin{gather}\label{T3}
\partial_i \rho^{(\ell)}_j+\big(\rho^{(\ell)}_j-\rho^{(\ell)}_i\big)\frac{\partial_i H^{-2}_j}{H^{-2}_j}+\rho^{(N)}_j\sum_{h=1}^N\big(\rho^{(\ell)}_h-\rho^{(\ell)}_i\big) H^{-2}_h\partial_ia_h=0.
\end{gather}
Note that if $\partial_i a_h=0$ for $i\ne h$ then we recover St\"ackel form and (\ref{T3}) simplif\/ies to Eisenhart's equation~\cite{EIS48, EIS34}
\begin{gather*}
\partial_i \rho^{(\ell)}_j+\big(\rho^{(\ell)}_j-\rho^{(\ell)}_i\big)\frac{\partial_i H^{-2}_j}{H^{-2}_j}=0.
\end{gather*}

A way of expressing the identity (\ref{T3}) that does not require the introduction of the terms $a_h$ is to note that at least one of the $\rho^{(N)}_j$, must be nonzero, say for $j=1$.
Setting $j=1$ in (\ref{T3}) we obtain
\[
\sum_{h=1}^N \big(\rho^{(\ell)}_h-\rho^{(\ell)}_i\big)  H^{-2}_h\partial_ia_h=-\frac{1}{\rho^{(N)}_1} \left(\partial_i \rho^{(\ell)}_1+\big(\rho^{(\ell)}_1-\rho^{(\ell)}_i\big)\frac{\partial_i H^{-2}_1}{H^{-2}_1}\right).
\]
Substituting this result back into (\ref{T3}) we conclude that
\begin{gather}\label{T5}
\partial_{i}\rho^{(\ell)}_j+\big(\rho^{(\ell)}_j-\rho^{(\ell)}_i\big)\frac{\partial_i  H^{-2}_j}{H^{-2}_j}=
\frac{\rho^{(N)}_j}{\rho^{(N)}_1}\left(\partial_{i}\rho^{(\ell)}_1+\big(\rho^{(\ell)}_1-\rho^{(\ell)}_i\big)\frac{\partial_i  H^{-2}_1}{H^{-2}_1}\right).
\end{gather}

For future use we remark that if we restrict to the case $\ell=N$ then, for $B_j=\rho^{(N)}_j/\rho^{(N)}_1$, (\ref{T5})~becomes
\begin{gather}\label{T6}
\partial_{i}B_j=
(B_i-B_j)\frac{\partial_{i} H_j^{-2}}{H^{-2}_j} + B_j(1-B_i)\frac{\partial_{i}H_1^{-2}}{H^{-2}_1},\qquad i,j=1,\dots,N.
\end{gather}

\begin{remark}  There is an equivalence relation obeyed  by generalized St\"ackel matrices. If $S$ is the matrix  (\ref{genstackel1}), then for any nonzero function $f({\bf x})$, the generalized St\"ackel matrix
\begin{gather}\label{genstackelequiv}
S'=  \begin{pmatrix} s_{11}\big(x^1\big) & s_{12}\big(x^1\big) & \cdots &s_{1,N-1}\big(x^1\big) & a_1({\bf x})f({\bf x})\\
s_{21}\big(x^2\big) & s_{22}\big(x^2\big) & \cdots &s_{2,N-1}\big(x^2\big) & a_2({\bf x})f({\bf x})\\
\cdots &\cdots &\cdots &\cdots &\cdots\\
s_{N1}\big(x^N\big) & s_{N2}\big(x^N\big) & \cdots &s_{N,N-1}\big(x^N\big) & a_N({\bf x})f({\bf x})\end{pmatrix},
\end{gather}
def\/ines exactly the same Hamilton--Jacobi equation, separation equations  and side condition as does~$S$.
\end{remark}

\begin{remark}
 This construction of Hamilton--Jacobi systems  with a side condition can easily be extended  to construct systems with two or more side conditions. For example, with  two side conditions
${\cal L}^{N-1}=0$, ${\cal L}^N=0$, the last two columns
of the $N\times N$ generalized St\"ackel matrix would be arbitrary and the  symmetries  would be modulo the side conditions:
\[
\big\{{\cal L}^i,{\cal L}^j\big\}=A_{i,j}{\cal L}^{{N}}+B_{i,j}{\cal L}^{N-1},\qquad i,j=1,\dots,N.
\]
In a similar fashion nonregular separability of Helmholtz equations with multiple linear side conditions can be def\/ined.
\end{remark}

\section{Generalized St\"ackel form for the Helmholtz equation}\label{Helmholtzstackel}

Now we perform an analogous construction of eigenfunctions  for a Helmholtz-like operator, using the same generalized St\"ackel matrix~$S$. We want eigenfunctions of the separated form $\Psi=\prod\limits_{j=1}^N\Psi^{(j)}(x^j)$. We take separation equations in the form
\begin{gather*}
\partial_\ell^2\Psi+f_\ell\big(x^\ell\big)\partial_\ell\Psi+\left(v_\ell\big(x^\ell\big)-\sum_{\alpha=1}^{N-1}s_{\ell \alpha}\big(x^\ell\big)\lambda_\alpha\right)\Psi=0,\qquad \ell=1,\dots,N.
\end{gather*}
Then we have the eigenvalue equations
\begin{gather*}
L^\beta\Psi\equiv \sum_{\ell=1}^NT^{\beta\ell}\big(\partial_\ell^2+f_\ell\ \partial_\ell+v_\ell\big)\Psi=\lambda_\beta\Psi,\qquad \beta=1,\dots,N-1,
\end{gather*}
and the side condition
\begin{gather*}
L^N\Psi\equiv  \sum_{\ell=1}^NT^{N\ell}\big(\partial_\ell^2+f_\ell\partial_\ell+v_\ell\big)\Psi=0.
\end{gather*}
We take $L^1=H$, $\lambda_1=E$, so ${-\frac{1}{2}}H$ is the { standard} Hamiltonian operator.

Let
\[
X_\ell=\partial_\ell^2+f_\ell\ \partial_\ell+v_\ell,\qquad Y_\ell=\partial_\ell^2+f_\ell\ \partial_\ell.
\]
 We need to compute the commutator $[L^\alpha,L^j]$ for $\alpha=1,\dots,N-1$, $j=1,\dots,N$. Now
\begin{gather*}
L^\alpha L^j= \left(\sum_i\rho^{(\alpha)}_iH_i^{-2}X_i\right)
\left(\sum_k\rho^{(j)}_k H_k^{-2}X_k\right)=\sum_{i,k}\rho^{(\alpha)}_i\rho^{(j)}_k H_i^{-2}H_k^{-2}X_iX_k\\
\hphantom{L^\alpha L^j=}{}
+\sum_{i,k}\rho^{(\alpha)}_i H_i^{-2}Y_i\big(\rho^{(j)}_kH_k^{-2}\big)X_k+2\sum_{i,k}\rho^{(\alpha)}_i H_i^{-2}\partial_i\big(\rho^{(j)}_kH_k^{-2}\big)\partial_iX_k,\\
L^jL^\alpha= \left(\sum_k\rho^{(j)}_k H_k^{-2}X_k\right)
\left(\sum_i\rho^{(\alpha)}_iH_i^{-2}X_i\right)=\sum_{i,k}\rho^{(j)}_k\rho^{(\alpha)}_i H_k^{-2}H_i^{-2}X_kX_i\\
\hphantom{L^jL^\alpha=}{}
+\sum_{i,k}\rho^{(j)}_k H_k^{-2}Y_k\big(\rho^{(\alpha)}_iH_i^{-2}\big)X_i+2\sum_{i,k}\rho^{(j)}_k H_k^{-2}\partial_k\big(\rho^{(\alpha)}_iH_i^{-2}\big)\partial_kX_i,
\end{gather*}
so
\begin{gather*}
\big[L^\alpha,L^j\big]=\sum_{i,k}\left(\rho^{(\alpha)}_iY_i\big(\rho^{(j)}_kH_k^{-2}\big)-\rho_i^{(j)}Y_i\big(\rho_k^{(\alpha)}H_k^{-2}\big)   \right) H_i^{-2}X_k\\
\hphantom{\big[L^\alpha,L^j\big]=}{}
+2\sum_{i,k}\left(\rho^{(\alpha)}_i \partial_i\big(\rho^{(j)}_kH_k^{-2}\big)-\rho^{(j)}_i \partial_i\big(\rho^{(\alpha)}_kH_k^{-2}\big)\right)H_i^{-2}\partial_iX_k.
\end{gather*}
Using (\ref{T3}) we can establish the identities
\begin{gather*} \left(\rho^{(\alpha)}_i \partial_i\big(\rho^{(j)}_kH_k^{-2}\big)-\rho^{(j)}_i \partial_i\big(\rho^{(\alpha)}_kH_k^{-2}\big)\right)H_i^{-2}
\partial_i\nonumber\\
\qquad{} = \rho_k^{(N)}H_k^{-2}H_i^{-2}\sum_{h=1}^N\big(\rho^{(\alpha)}_h\rho^{(j)}_i-\rho^{(\alpha)}_i\rho^{(j)}_h\big)H_h^{-2}(\partial_i a_h)\partial_i,\\
\rho_k^{(N)}H_k^{-2}\partial_iF=-\left(\!\rho^{(N)}_i\partial_i H_k^{-2}+\rho^{(N)}_kH_k^{-2}\sum_{h=1}^N\big(\rho^{(N)}_i-\rho^{(N)}_h\big)H^{-2}_h\partial_i a_h\!\right)\!F
+\partial_i\big(\rho_k^{(N)}H_k^{-2}F\big),
\end{gather*}
for any function $F$, and
\begin{gather*}
\left(\rho^{(\alpha)}_iY_i\big(\rho^{(j)}_kH_k^{-2}\big)-\rho_i^{(j)}Y_i\big(\rho_k^{(\alpha)}H_k^{-2}\big)   \right) H_i^{-2}\nonumber\\
\qquad{}{}
=2\rho_i^{(N)}H_i^{-2}\partial_iH_k^{-2}\sum_{h=1}^N\big(\rho^{(\alpha)}_h\rho^{(j)}_i-\rho^{(j)}_h\rho^{(\alpha)}_i\big)H_h^{-2}\partial_i a_h
\nonumber\\
\qquad\quad{} + \rho^{(N)}_kH_k^{-2}H_i^{-2}\left(\rho^{(j)}_i\partial_i\left(\sum_{h=1}^N\rho^{(\alpha)}_h H_h^{-2}\partial_i a_h\right)-\rho^{(\alpha)}_i\partial_i\left(\sum_{h=1}^N\rho^{(j)}_hH_h^{-2}\partial_i a_h\right)\right)\\
\qquad\quad{}
+\rho^{(N)}_kH_k^{-2}H_i^{-2}\left(\sum_{h=1}^N\big(\rho_i^{(N)}-\rho_h^{(N)}\big)H_h^{-2}\partial_i a_h\right) \left(\sum_{h=1}^N\big(\rho_i^{(j)}\rho_h^{(\alpha)}-\rho_i^{(\alpha)}\rho_h^{(j)}\big)H_h^{-2}\partial_i a_h\right)\!.\nonumber
\end{gather*}
Thus,
\begin{gather*}
\big[L^\alpha,L^j\big]=\sum_{i=1}^N H_i^{-2}\left(\left( 2\sum_{h=1}^N\big(\rho_i^{(j)}\rho_h^{(\alpha)}-\rho_i^{(\alpha)}\rho_h^{(j)}\big)H_h^{-2}\partial_i a_h\right)\right.\nonumber\\
\hphantom{\big[L^\alpha,L^j\big]=}{}
\times\left(-\frac12\sum_{h=1}^N\big(\rho_i^{(N)}-\rho_h^{(N)}\big)H_h^{-2}\partial_i a_h+\partial_i\right)
+\left(\sum_{h=1}^N\big(\rho_i^{(N)}-\rho_h^{(N)}\big)H_h^{-2}\partial_i a_h\right) \nonumber\\
\left. \hphantom{\big[L^\alpha,L^j\big]=}{}
\times\left(\sum_{h=1}^N\big(\rho_i^{(j)}\rho_h^{(\alpha)}-\rho_i^{(\alpha)}\rho_h^{(j)}\big)H_h^{-2}\partial_i a_h\right)\right)L^N
=F_{\alpha j}L^N,
\end{gather*}
where $F_{\alpha j}$ is a f\/irst-order partial dif\/ferential operator.

We see that  there is no obstruction to lifting  our classical nonregular separation to the operator case. A dif\/f\/iculty occurs, however, when we try to write the pure operator part of~$H$ as a~Laplace--Beltrami operator on a Riemannian manifold. Then there is an obstruction, a~gene\-ra\-li\-zed Robertson condition, to be worked out. Also we need to examine the ef\/fect of permitting $R$-separation.

\section{Maximal nonregular separation as regular separation\\ with a side condition} \label{nonregsep1}

Another way to approach the classical Hamilton--Jacobi  problem is to use the Kalnins--Miller method for variable separation~\cite{EGKWM83,EGKWM83+} and consider
maximal nonregular separation as regular separation with a side condition. (Here the nonregular separation is {\it maximal} in the sense that with a single side
condition  the number of separation constants is the maximum possible for nonregular separation, i.e., just 1 less than that for regular
separation.)  We look for additively separable solutions of the equation
\begin{gather} \label{I}
\sum_{i=1}^N H_i^{-2}u_i^2+V=E
\end{gather}
with the side condition
\begin{gather}\label{II} \sum_{i=1}^N L_i^{-2}u_i^2+W=0,\end{gather}
i.e., solutions $u$ such that $u_{ij}=0$ for $i\ne j$.
From (\ref{I}) we f\/ind
\[
 u_{jj}=-\frac{V_j+\sum_i\partial_j H_i^{-2}u_i^2}{2H_j^{-2}u_j},
\]
and from (\ref{II})
\[
 u_{jj}=-\frac{W_j+\sum_i\partial_j L_i^{-2}u_i^2}{2L_j^{-2}u_j}.
\]
These expressions must be equal modulo the side condition (\ref{II}), so
\begin{gather} \label{III}
\frac{V_j+\sum_i\partial_j H_i^{-2}u_i^2}{2H_j^{-2}u_j}=\frac{W_j+\sum_i\partial_j L_i^{-2}u_i^2}{2L_j^{-2}u_j}+{\nu}_j\left(\sum_iL_i^{-2}u_i^2+W\right),
\end{gather}
for some functions $\nu_j$.
Similarly, equations derived from $u_{jjk}=0$ for $j\ne k$ must hold modulo the side condition.
 Requiring that all of the above equations hold identically, i.e.,
requiring that we have regular separation modulo the side condition we  eventually obtain the conditions that
\begin{enumerate}\itemsep=0pt \item There are functions $\omega_{j\ell}=\omega_{\ell j}$ for all $j\ne \ell$ such that
\begin{gather}\label{i1}-\partial_{j\ell}L_i^{-2}+\frac{\partial_j L_\ell^{-2}}{L_\ell^{-2}}\partial_\ell L_i^{-2}+\frac{\partial_\ell L_j^{-2}}{L_j^{-2}}\partial_j L_i^{-2}+\omega_{j\ell}L_i^{-2}=0\end{gather}
and
\begin{gather}\label{i2}-W_{j\ell}+W_\ell\frac{\partial_j L_\ell^{-2}}{L_\ell^{-2}}+W_j\frac{\partial_\ell L_j^{-2}}{L_j^{-2}}+\omega_{j\ell}W=0\end{gather}
for all $i=1,\dots,N$. (If all the $L_\ell$ are nonzero, this  means that the  $L_i^{-2}$ are in conformal St\"ackel form, i.e., an arbitrary function times a St\"ackel form matrix,
and that $W$ is a~conformal St\"ackel form potential \cite{MIL83+,MIL83}.)
 \item There are functions $\tau_j$, $j=1,\dots,N$ such that
\begin{gather}\label{ii1}
\frac{\partial_jH_i^{-2}}{H_j^{-2}}=\frac{\partial_jL_i^{-2}}{ L_j^{-2}}+\tau _jL_i^{-2},
\end{gather}
for all $i=1,\dots,N$, and
\begin{gather}\label{ii2}
\frac{V_j}{H_j^{-2}}=\frac{W_j}{L_j^{-2}}+\tau_j W.
\end{gather}
\item Let $C_{ij}$ be the second-order dif\/ferential operator  acting on functions $f$ by
\begin{gather}\label{Cij}
C_{ij}(f)=\partial_{ij}f-\frac{\partial_j H_i^{-2}}{H_i^{-2}}\partial_i f-\frac{\partial_i\ln H_j^{-2}}{H_j^{-2}}\partial_j f.
\end{gather}
There are functions $\mu_{j\ell}=\mu_{\ell j}$ for all $j\ne \ell$ such that
\begin{gather}\label{iii1}
C_{j\ell}\big(H_i^{-2}\big)=\mu_{j\ell}L_i^{-2},
\end{gather}
and
\begin{gather}\label{iii2}
C_{j\ell}(V)=\mu_{j\ell}W
\end{gather}
for all $j,\ell=1,\dots,N$, with  $j\ne \ell$.
\end{enumerate}

\section[Maximal nonregular separation $\Rightarrow$ generalized St\"ackel form]{Maximal nonregular separation $\boldsymbol{\Rightarrow}$ generalized St\"ackel form}\label{nonregsep2}

We have  shown  that the f\/irst row of the inverse of a generalized St\"ackel matrix with arbitrary $N$-th column is an orthogonal metric $H_i^{-2}$
whose associated Hamilton--Jacobi equation admits nonregular separation on the hypersurfaces given by the level set $\mathcal{L}^N=0$ of a
 function $\mathcal{L}^N$,  quadratic in momentum variables, which is a constrained f\/irst integral on the same level set and whose components are the $N$-th row of the inverse of the
 generalized St\"ackel matrix.
Now we prove the  converse, i.e.,  that if an orthogonal geodesic Hamiltonian is separable in orthogonal coordinates on the level set of a
quadratic f\/irst integral, then it is a row of the inverse of a~gene\-ra\-li\-zed metric and the quadratic coef\/f\/icients of $\mathcal{L}^N$ are the $N$-th row of the generalized
St\"ackel matrix.
Our starting point here is the geometrical framework of regular separation of variables: a~complete separated solution of the Hamilton--Jacobi equation  is a foliation parametrized
by~$N$ parameters  for the integral manifold of the distribution generated by the $N$ vector f\/ields $D_i=\partial_{x^i} + R_i\partial_{y_i}$ where the $R_i$ are
determined by the condition that the $D_i$ are tangent to ${\cal H}= \operatorname{const}$. The classical Levi-Civita conditions~\cite{LC} represent the integrability conditions of the
 distribution.

In our case we need  this distribution  to be integrable  only on the submanifold $S$ def\/ined by $\mathcal{L}^N=0$. We also  need
 the vector f\/ields to be tangent to
the submanifold~$S$
(closely related  to the compatibility of the side condition). Our  f\/irst step will be to write the dif\/ferential conditions that in this case play the role of the Levi-Civita condition for regular separation.
In this case they mix $\cal H$ and $\cal{L}^N$. Moreover, they include also the condition that $D_i$ are tangent to $\mathcal{L}^N=0$. We will show that these equations are exactly
the equations for nonregular separation derived in Section~\ref{nonregsep1}.
Then we will relate these conditions to the existence of some more intrinsic geometrical object
for the eigenvalues of  Killing tensors (associated with quadratic in the momenta constants of motion), whose integrability conditions are equivalent to the Levi-Civita
conditions.

Finally,  we will construct the family of quadratic ``f\/irst integrals'' $\mathcal{L}^h=\sum\limits_{j=1}^N T^{hj}u_{j}^2$ commuting and constant for the motion on $\mathcal{L}^N=0$
diagonalized
in the coordinates we are considering,
and show that the inverse of the matrix of the components $T^{hj}$ is a generalized St\"ackel matrix. Then we will extend our analysis to prove the corresponding
results for multiplicative separation or $R$-separation of the Helmholtz
or time independent Schr\"odinger equations.

\subsection{Dif\/ferential conditions for nonregular separation\\ on a quadratic f\/irst integral leaf }\label{diffconds}
As in Section~\ref{nonregsep1} we  consider a natural Hamiltonian in orthogonal coordinates $\mathbf{x}=(x^i)$ on the
cotangent bundle of a $N$-dimensional Riemannian manifold $Q$ with Hamiltonian
\begin{gather} \label{hort}
\mathcal{H} = \mathcal{L}^1 = \sum_{i=1}^{N}H_i^{-2}p_i^2 + V(\mathbf{x}),
\end{gather}
and a function $\mathcal{L}^N$ which is also quadratic in the momenta $(p_i)$ and diagonalized
in the same coordinates
\[
\mathcal{L}^N =
 \sum_{i=1}^{N}\rho^{(N)}_iH_i^{-2}p_i^2 + W(\mathbf{x}),
\]
where $\rho^{(N)}_i$ are the eigenvalues with respect to the metric $H^{-2}_i$, i.e., $L_i=\rho^{(N)}_iH_i^{-2}$.
We want to study the existence of separated solutions $u$ of the Hamilton--Jacobi
equation
\begin{gather}\label{HJ1}
 \sum_{i=1}^{N}H_i^{-2}u_i^2 + V(\mathbf{x})=E, \qquad u_i=\partial_{i}u,
\end{gather}
with the side condition, or constraint,
\[
 \sum_{i=1}^{N}\rho^{(N)}_iH_i^{-2}u_i^2 + W(\mathbf{x})=0.
\]
Recall that the conditions for nonregular separation are (\ref{i1}), (\ref{i2}), (\ref{ii1}), (\ref{ii2}), (\ref{iii1}), (\ref{iii2}) for $L_i=\rho_i^{(N)}H_i^{-2}$ and
some functions $\omega_{\ell j}$, $\tau_j$, $\mu_{j\ell}$.
However, in this case,  we can avoid the introduction of additional unknown functions, since we can solve the equation $\mathcal{L}^N = 0$ with
respect to a~momentum variable $u_i$ (as for instance   $u_1^2=u_1^2(x^j,u_\alpha)$ with $\alpha=2,\ldots,N$)
so an expression vanishes on $\mathcal{L}^N=0$ if and only if it vanishes for all $u_\alpha$ for $\alpha=2,\ldots,N$ after the substitution of~$u_1$ by ${u_1}(x^j,u_\alpha)$.
This simplif\/ies very much the task of f\/inding equivalent conditions such as the link with generalized St\"ackel matrices and it is
possible only because we are assuming orthogonal coordinates.

The conditions  (\ref{i1}), (\ref{i2}), (\ref{ii1}), (\ref{ii2}), (\ref{iii1}), (\ref{iii2}) are equivalent (supposing without loss of generality that $\rho^{(N)}_1\neq 0$)
 to imposing that  for
\begin{gather}\label{LN0}
u_1^2 = - \frac{W}{\rho^{(N)}_1H_1^{-2}}- \sum_{\alpha=2}^{N}\frac{\rho^{(N)}_\alpha H_\alpha^{-2}}{\rho^{(N)}_1H_1^{-2}} u_\alpha^2
\end{gather}
the expressions (\ref{i1})--(\ref{iii2}) vanish for all values of $u_\alpha^2$.
Even easier, inserting (\ref{LN0}) in (\ref{III}) we f\/ind that the coef\/f\/icient of $\nu_j$ vanishes and equating coef\/f\/icients  of $u_\alpha^2$  we get
\begin{gather}
\frac{\partial_{j}\rho^{(N)}_\alpha H_\alpha^{-2}}{\rho^{(N)}_\alpha H_\alpha^{-2}} - \frac{\rho^{(N)}_j}{\rho^{(N)}_\alpha} \frac{\partial_{j} H_\alpha^{-2}}{H_\alpha^{-2}}
 =\frac{\partial_{j}\rho^{(N)}_1 H_1^{-2}}{\rho^{(N)}_1 H_1^{-2}} - \frac{\rho^{(N)}_j}{\rho^{(N)}_1} \frac{\partial_{j} H_1^{-2}}{H_1^{-2}}
 =\frac{\partial_{j}W}{W} - \frac{\rho^{(N)}_j}{W} \partial_{j}V. \label{EiLN0}
\end{gather}
 Similarly we f\/ind
\begin{gather}
\label{LCLN0}
\frac{C_{ij}\big(H_\alpha^{-2}\big)}{\rho^{(N)}_\alpha H_\alpha^{-2}}=\frac{C_{ij}\big(H_1^{-2}\big)}{\rho^{(N)}_1 H_1^{-2}}=
\frac{C_{ij}(V)}{W}.
\end{gather}
In all these equations we follow the convention that the vanishing of a factor $\rho^{(N)}_\alpha$ or $W$ in a~denominator in one of  expressions (\ref{EiLN0}), (\ref{LCLN0}), implies that
the numerator vanishes.

Note that (\ref{LCLN0}) and (\ref{EiLN0}) give conditions for separation of  the geodesic Hamiltonian with $V=W=0$, with additional conditions that $V$ and $W$ must satisfy.
Supposing for simplicity for the moment $V=W=0$,  we can rewrite
(\ref{LCLN0}) and (\ref{EiLN0}) as
\begin{gather}
\label{LCLN}
\frac{C_{ij}\big(H_\alpha^{-2}\big)}{ H_\alpha^{-2}}=\frac{\rho^{(N)}_\alpha}{\rho^{(N)}_1}\frac{C_{ij}\big(H_1^{-2}\big)}{ H_1^{-2}},\\
 \label{EiLN}
\partial_{i}\frac{\rho^{(N)}_\alpha}{\rho^{(N)}_1 }=
\left(\frac{\rho^{(N)}_i}{\rho^{(N)}_1 }-\frac{\rho^{(N)}_\alpha}{\rho^{(N)}_1 }\right)
\frac{\partial_{i} H_\alpha^{-2}}{H_\alpha^{-2}}
+ \frac{\rho^{(N)}_\alpha}{\rho^{(N)}_1 }\left(1-\frac{\rho^{(N)}_i}{\rho^{(N)}_1 }\right)\frac{\partial_{i}H_1^{-2}}{H_1^{-2}},
\end{gather}
which are the necessary and suf\/f\/icient conditions.
Equations (\ref{EiLN}) can be interpreted as a f\/irst-order system in the $N-1$ unknowns $B_\alpha= \frac{\rho^{(N)}_\alpha}{\rho^{(N)}_1}$.
\begin{proposition}
A geodesic Hamiltonian $\cal H$ admits nonregular separation  on the submanifold $\mathcal{L}^N=0$ in a given orthogonal coordinate system if and only if
the functions $B_k= \frac{\rho^{(N)}_k}{\rho^{(N)}_1}$ $(k=1,\ldots,N)$ satisfy
\begin{gather}
\label{EiLN1}
\partial_{i}B_k=
(B_i-B_k)\frac{\partial_{i} H_k^{-2}}{H_k^{-2}} + B_k(1-B_i)\frac{\partial_{i}H_1^{-2}}{H_1^{-2}}.
\end{gather}
\end{proposition}

\begin{proof}
By rewriting the necessary and suf\/f\/icient conditions (\ref{LCLN})
and (\ref{EiLN}) in terms of the  $N$ functions $B_k=B_1$, $B_\alpha$ with $B_1=1$ we get
\begin{gather}
 \label{LCLN1}
\frac{C_{ij}\big(H_k^{-2}\big)}{ H_k^{-2}}=B_k\frac{C_{ij}\big(H_1^{-2}\big)}{ H_1^{-2}}
\end{gather}
and (\ref{EiLN1}). However, it is a straightforward calculation that (\ref{LCLN1}) is a dif\/ferential consequence of~(\ref{EiLN1}), indeed we have
\begin{gather}\label{intcondLN}
\partial_i \partial_j B_k - \partial_j \partial_i B_k= (B_i-B_j)\left(\frac{C_{ij}\big(H_k^{-2}\big)}{H_k^{-2}}
-B_k\frac{C_{ij}\big(H_1^{-2}\big)}{ H_1^{-2}} \right).\qquad {\qed}
\end{gather}
\renewcommand{\qed}{}
\end{proof}

Note that the system of equations (\ref{EiLN1}) coincides with (\ref{T6}), verifying again that separation with a generalized St\"ackel matrix is nonregular.

\begin{proposition}
A natural Hamiltonian $\cal H$ admits nonregular separation  on the submanifold $\mathcal{L}^N=0$ in a given orthogonal coordinate system only if the ratios
\[
\frac{C_{ij}(H_\alpha^{-2}) H_1^{-2}}{C_{ij}(H_1^{-2}) H_\alpha^{-2}}
\]
are independent of $i$ and $j$ and the eigenvalues of the quadratic function $\mathcal{L}^N$ are proportional to them
\begin{gather}\label{LNc}
B_\alpha=\frac{\rho^{(N)}_\alpha}{\rho^{(N)}_1}= \frac{C_{ij}\big(H_\alpha^{-2}\big) H_1^{-2}}{C_{ij}\big(H_1^{-2}\big) H_\alpha^{-2}}.
\end{gather}
\end{proposition}

\begin{remark}
The function $\mathcal{L}^N$ is naturally def\/ined up to a multiplicative factor $f$ on $Q$: if an expression is zero on  $\mathcal{L}^N=0$ then it is also
zero on $f\mathcal{L}^N=0$ for all functions on $T^*Q$, but, since we are interested on quadratic in the momenta functions,  we can normalize $f$ on $Q$.
(From the St\"ackel matrix point of view this corresponds to
the multiplication of the $N$-th column by~$f$.
Hence, equations (\ref{LCLN1}) determine the unique (up to a factor)  quadratic hypersurface where separation could occur.
Indeed, equations (\ref{intcondLN}) are the complete integrability conditions for the f\/irst-order PDE system~(\ref{EiLN1}).
These conditions are identically satisf\/ied for all~$(B_k, x^i)$ on an open subset of $\mathbb C^{2n}$ only if $C_{ij}(H_k^{-2})=0$, that is only if regular
 separation occurs.
However, a single solution $B_k=B_k(x^i)$ could exist, provided it satisf\/ies the original conditions (\ref{EiLN1}), that is if it takes the form
(\ref{LNc}), or with all $B_k$ equal, that is for $\mathcal{L}^N={\cal H}$. In this case we get separation of the null equation
${\cal H}=0$ (in which  case a column of the St\"ackel matrix was arbitrary, so the result  is consistent with the known one).
\end{remark}

Let us examine the relation of this kind of separation with the generalized St\"ackel matrix.

\begin{theorem}\label{theorem1}
Suppose  the natural Hamiltonian \eqref{hort} admits  $N-2$ other functions $\mathcal{L}^2,\ldots,\mathcal{L}^{N-1}$, quadratic in the momenta
 such that
\begin{enumerate}\itemsep=0pt
\item[$1)$] $\mathcal{L}^1={\cal H},\mathcal{L}^2,\ldots,\mathcal{L}^{N-1}, \mathcal{L}^N$ are pointwise independent,
\item[$2)$] $\mathcal{L}^1={\cal H},\mathcal{L}^2,\ldots,\mathcal{L}^{N-1}, \mathcal{L}^N$ are constants of the motion, modulo $\mathcal{L}^N$,
that is
\[
\big\{{\cal H},\mathcal{L}^k\big\}|_{\mathcal{L}^N=0}=0,
\]
\item[$3)$] the quadratic terms of $\mathcal{L}^1={\cal H},\mathcal{L}^2,\ldots,\mathcal{L}^{N-1}, \mathcal{L}^N$ are diagonal in the coordinates $(x^i)$, that is
\[
\mathcal{L}^\ell= \sum_{i=1}^N \rho^{(\ell)}_i H_i^{-2}\big( u_i^2+v_i({\bf x})\big), \qquad \rho^{(1)}_i=1\ \ \forall\; i.
\]
\end{enumerate}
Then,
\begin{enumerate}\itemsep=0pt
\item[$1)$] \label{i} $\partial_jv_i=0$ for $j\ne i$ and
the eigenvalues $\rho^{(\ell)}_i$ of $\mathcal{L}^\ell$ satisfy the following generalization of the Eisenhart conditions
\begin{gather}\label{Eisgen}
\partial_{i}\rho^{(\ell)}_\alpha+\big(\rho^{(\ell)}_\alpha-\rho^{(\ell)}_i\big)\frac{\partial_i H^{-2}_\alpha}{H^{-2}_\alpha}=
\frac{\rho^{(N)}_\alpha}{\rho^{(N)}_1}\left(\partial_{i}\rho^{(\ell)}_1+\big(\rho^{(\ell)}_1-\rho^{(\ell)}_i\big)\frac{\partial_i H^{-2}_1}{H^{-2}_1}\right);
\end{gather}
\item[$2)$]\label{ii}
the inverse matrix of $T^{\ell i}=\big(\rho^{(\ell)}_i H_i^{-2}\big)$ is a generalized St\"ackel matrix with the last column made of arbitrary functions of $N$ variables $a_1(\mathbf x),\ldots, a_n(\mathbf x)$;
\item[$3)$] \label{iii}
the Hamilton Jacobi equation \eqref{HJ1} admits a maximal nonregular separated solution on $\mathcal{L}^N=0$ depending on $N-1$ parameters;
\item[$4)$] \label{iv}
the $N$ functions $\mathcal{L}^1={\cal H},\mathcal{L}^2,\ldots,\mathcal{L}^{N-1}, \mathcal{L}^N$ are in involution on $S$,
that is
\[
\big\{\mathcal{L}^\ell,\mathcal{L}^k\big\}|_{\mathcal{L}^N=0}.
\]
\end{enumerate}
\end{theorem}

\begin{proof}
1)
By inserting (\ref{LN0}) in the Poisson brackets
\begin{gather*}
\big\{\mathcal{L}^\ell,{\cal H}\big\}= \sum_{i}2H_i^{-2}u_i\left(\sum_{k=1}^N \big(  \partial_{i} \big(\rho^{(\ell)}_k H_k^{-2}\big) -
 \rho^{(\ell)}_i \partial_{i} H_k^{-2}\big)u_k^2 \right.\\
\left.
\hphantom{\big\{\mathcal{L}^\ell,{\cal H}\big\}=}{}
+\sum_{k=1}^N\big( \partial_{i}\big(\rho_k^{(\ell)}H_k^{-2}v_k\big)-\rho_i^{(\ell)}\partial_i\big(H_k^{-2}v_k\big)\big)\right),
\end{gather*}
we get that $\partial_iv_k=0$ for $i\ne k$ and that $\rho^{(\ell)}_k$ satisf\/ies the generalized Eisenhart conditions (\ref{Eisgen}).

2) Since the quadratic diagonal functions are  pointwise independent, the matrix $T^{\ell i}$ has a~nonzero determinant and therefore it admits an inverse $s_{kh}$ such that
\begin{gather} \label{inv}
 T^{\ell k} s_{kh}=\delta^\ell_h
\end{gather}
(we partially follow the proof of equation (\ref{T3}) and a proof of the Eisenhart theorem given in~\cite{DegRast}).
We want to show that $\partial_i s_{h\xi}=0$ for all $i\neq h$ and $\xi\neq N$ if and only if $\rho_j^{(\ell)}=T^{\ell j}/T^{1 j}$ satisfy the generalized
Eisenhart conditions (\ref{Eisgen}). Dif\/ferentiating (\ref{inv}) with respect to $x^i$ we get
\begin{gather*}
 \sum_k \big(\partial_i T^{\ell k} s_{kh} +  T^{\ell k} \partial_i s_{kh}\big)=0,
\end{gather*}
that is
\begin{gather}
 \label{e1}
 \sum_k \big(\big( H_k^{-2}\partial_i \rho_k^{(\ell)} +  \rho_k^{(\ell)}\partial_i H_k^{-2} \big)s_{kh} +  \rho_k^{(\ell)} H_k^{-2} \partial_i s_{kh}\big)=0, \qquad   \ell\neq 1,
\\
\label{e2}
 \sum_k \big(\partial_i H_k^{-2} s_{kh}+ H_k^{-2} \partial_i s_{kh}\big)=0, \qquad  \ell= 1.
\end{gather}
Adding (\ref{e2}) multiplied by $-\rho^{(\ell)}_i$ to equation  (\ref{e1}) we get
\begin{gather}\label{e3}
 \sum_k \big(\big(H_k^{-2}\partial_i \rho_k^{(\ell)}  +  \big(\rho_k^{(\ell)}-\rho_i^{\ell}\big)\partial_i H_k^{-2} \big)s_{kh} +
\big(\rho_k^{(\ell)}-\rho_i^{(\ell)}\big) H_k^{-2} \partial_i s_{kh}\big)=0, \qquad \ell\neq 1.
\end{gather}
Let us suppose that $\partial_i s_{kh}=0$ for all $i\neq k$ and $h\neq N$ and $\partial_i s_{kN}=\partial_i a_h$, by multiplying
(\ref{e3}) by $T^{hj}$, we get
\[
\sum_k \big(H_k^{-2}\partial_i \rho_k^{(\ell)}  +  \big(\rho_k^{(\ell)}-\rho_i^{(\ell)}\big)\partial_i H_k^{-2}\big)\delta^k_j +
 \sum_h \sum_k\big(\big(\rho_k^{(\ell)}-\rho_i^{(\ell)}\big) H_k^{-2} \partial_i s_{kh}\big)T^{hj}=0,
\]
that is
\begin{gather}\label{e5}
 H_j^{-2}\partial_i \rho_j^{(\ell)}  +  \big(\rho_j^{(\ell)}-\rho_i^{(\ell)}\big)\partial_i H_j^{-2} +
T^{Nj}  \sum_k\big(\big(\rho_k^{(\ell)}-\rho_i^{(\ell)}\big) H_k^{-2} \partial_i a_k\big)=0.
\end{gather}
Writing (\ref{e5}) for $j=1$ and $j=\alpha$ and by combining them in order to eliminate the common  factor
$ \sum_k\big(\big(\rho_k^{(\ell)}-\rho_i^{(\ell)}\big) H_k^{-2} \partial_i a_k\big)$, we get (\ref{Eisgen}).
 Conversely, let us suppose that (\ref{Eisgen}) hold; this means that the quantities
\[
\frac{H_k^{-2}\partial_i \rho_k^{(\ell)}  +  \big(\rho_k^{(\ell)}-\rho_k^{(\ell)}\big)\partial_i H_k^{-2}}{T^{Nk}}
\]
are independent of $k$, that is $H_k^{-2}\partial_i \rho_k^{(\ell)}  +  \big(\rho_k^{(\ell)}-\rho_k^{(\ell)}\big)\partial_i H_k^{-2}=T^{Nk} Q^{\ell}_i$.
Therefore (\ref{e3}) becomes
\[
T^{Nk} Q^{\ell}_i s_{kh} +
 \sum_k \big(\big(\rho_k^{(\ell)}-\rho_i^{(\ell)}\big) H_k^{-2} \partial_i s_{kh}\big)=
\delta_h^N Q^{\ell}_i+ \sum_k \big(\big(\rho_k^{(\ell)}-\rho_i^{(\ell)}\big) H_k^{-2} \partial_i s_{kh}\big)=0.
\]
If we f\/ix the values of $h$ and $i$, then  the above equations are a linear system of $N-1$ equations ($\ell=2,\ldots,N$)
in $N-1$ unknown $\partial_i s_{kh}$ $k=1,\ldots,N$, $k\neq h$). For $h\neq N$ the system is homogeneous and the matrix of coef\/f\/icient
$M^{\ell k}=\big(\rho_k^{(\ell)}-\rho_i^{(\ell)}\big)H_k^{-2}$ has nonzero determinant (see~\cite{DegRast}). Hence, the only  solution is $\partial_i s_{kh}=0$ for $h\neq N$.

3)~If equations (\ref{Eisgen}) hold, they hold in particular for $\ell=N$ and we get (\ref{EiLN}) which are proven to be equivalent to conditions
  (\ref{EiLN1}) that are necessary and suf\/f\/icient
for the nonregular separation with the side condition $\mathcal{L}^N=0$. Further, the conditions (\ref{EiLN0}) on $V$, $W$ are exactly the consistency conditions that
must be satisf\/ied by $V$, $W$ in order that $\{{\cal H},{\cal L}^N\}=0$ hold on ${\cal L}^N=0$, and conditions  (\ref{LCLN0}) are exactly the integrability conditions for
$V$, $W$ that must be satisf\/ied if $\{{\cal H},{\cal L}^N\}=0$ is to hold.

4)~This is essentially the computation in Section~\ref{genstac}.
\end{proof}

\begin{theorem}\label{theorem2}
If the $N$ pointwise independent functions
\[
\mathcal{L}^1={\cal H},\mathcal{L}^2,\ldots,\mathcal{L}^{N-1}, \mathcal{L}^N
\]
are in involution on $\mathcal{L}^N=0$ and admit common eigenvectors, i.e., if  the quadratic forms in the momenta associated with  ${\cal L}^\ell-\rho^{(\ell)}{\cal H}$ have $N$ common eigenvectors,  then the eigenvectors are normal:  there exist
orthogonal coordinates~$(x^i)$ such that the~$N$ functions are simultaneously diagonalized.
\end{theorem}

\begin{remark} This is essentially the converse of the results of Section~\ref{genstac}. It shows that separation with a side condition implies the existence of a generalized St\"ackel matrix that
def\/ines the separation.
\end{remark}

\begin{proof}
Apply Theorem~7.5 of \cite{HJ0} with $M_{ab}=K^N$, where $K^N$ is the symmetric 2-tensor associated with $\mathcal{L}^N$.
\end{proof}

\begin{remark}\label{remark5}
The functions $\mathcal{L}^\ell$ are def\/ined up to multiples of $\mathcal{L}^N$  and the conformal Killing tensors are def\/ined up to multiple of the metric tensor. Indeed, $\mathcal{L}^\ell+f^\ell\mathcal{L}^N$ have the same eigenvalues of~$\mathcal{L}^\ell$ (provided that the $\mathcal{L}^\ell$ have common eigenvectors) and they are in involution when restricted to $\mathcal{L}^N=0$:
\[
\big\{\mathcal{L}^\ell+f^\ell\mathcal{L}^N,\mathcal{L}^k+f^k\mathcal{L}^N\big\}={\cal Q}{\cal L}^N.
\]
\end{remark}

Now we close the logical loop. Suppose the conditions (\ref{EiLN1})--(\ref{intcondLN}) of maximal non\-re\-gu\-lar separation are satisf\/ied and consider
 the linear system of equations
\begin{gather}\label{T7}
\partial_{i}\rho_j+(\rho_j-\rho_i)\frac{\partial_i H^{-2}_j}{H^{-2}_j}=
B_j\left(\partial_{i}\rho_1+(\rho_1-\rho_i)\frac{\partial_i H^{-2}_1}{H^{-2}_1}\right),\qquad i,j=1,\dots,N,
\end{gather}
 for the unknowns $(\rho_1,\dots, \rho_N)$. Remark~\ref{remark5} suggests that this system has solutions for any choice of~$\rho_1$. Therefore, we choose the function $\rho_1$ arbitrarily and consider~(\ref{T7}) as a system of $N(N-1)$ independent dif\/ferential equations for the unknowns $(\rho_2,\dots, \rho_N)$.
It is then a straightforward  exercise to verify that the integrability conditions for~(\ref{T7}) are satisf\/ied identically,
due to the nonregular separation conditions. We already know the two solutions $(1,1,\dots,1)$ and $(B_2,\dots, B_N)$.  Indeed about any regular point ${\bf x}_0$ we can f\/ind a unique solution such that $\rho_j({\bf x}_0)=\beta_j$, $2\le j\le N$ for any choice of constants $\beta_j$. For example we could take $\rho_1\equiv 1$ and def\/ine $N-1$ linearly independent solutions by choosing the $\beta_j$ accordingly. Then we could take $\rho_1\equiv 0$ and f\/ind another solution such that the full $N$ solutions form a basis. This basis determines a generalized St\"ackel matrix.

It remains to consider the potential terms, the given functions $V$ and $W$. Using the functions~$\rho^{(\ell)}_i$ of the generalized St\"ackel matrix computed above we def\/ine functions
\[
\mathcal{L}^\ell= \sum_{i=1}^N \rho^{(\ell)}_i H_i^{-2} u_i^2 +W^{(\ell)}, \qquad \ell=1,\dots,N-1,
\]
where $V=W^{(1)}$ but the remaining $W^{(\ell)}$ will be determined by imposing the requirement that there  exist functions $ F_j$
linear in the momenta and such that
\[
\big\{{\cal H},{\cal L}^N\big\}=F_N{\cal L}^N,\qquad \big\{{\cal H},{\cal L}^\ell\big\}=F_\ell{\cal L}^N.
\]
By working out the Poisson brackets, it is straightforward to verify that equations~(\ref{EiLN0})
 are exactly the consistency conditions for the existence of~$F_N$ and~(\ref{LCLN0}) are exactly the consistency conditions for the  $F_\ell$ and the integrability conditions for the existence of the potentials~$W^{(\ell)}$.

 Now note that since the generalized St\"ackel matrix is invertible we can uniquely determine functions $v_i({\bf x})$ such that
\[
W^{(j)}=\sum_{i=1}^Nv_i\rho^{(j)}_iH_i^{-2},\qquad j=1,\dots, N.
\]
 With this observation we have verif\/ied all of the assumptions of Theorem~\ref{theorem1}. This proves the following result.

\begin{theorem}\label{theorem3}
If a  natural Hamiltonian $\cal H$ admits maximal nonregular separation  on the sub\-mani\-fold $\mathcal{L}^N=0$ in a given orthogonal coordinate system,
then the system is separable with a~side condition and there exists  a~generalized St\"ackel matrix for the separation.
\end{theorem}

All of these results for additive nonregular separation of Hamilton--Jacobi equations extend to multiplicative nonregular separation for the Helmholtz (or Schr\"odinger) equations on the same manifold, but with an obstruction. If a system is nonregular  $R$-separable  for the Helmholtz equation then it is nonregular separable for the Hamilton--Jacobi equation. However, if the Hamilton--Jacobi equation admits nonregular separation in some coordinate system then there is a ``generalized Robertson condition" to be solved to determine for which potentials the Helmholtz equation admits $R$-separation in these coordinates.  On the other hand, we can  always f\/ind some family of vector potentials for which the Helmholtz equation does admit nonregular $R$-separation in the coordinates.  That is, it is easy to show that given a nonregular separable system for the Hamilton--Jacobi equation, we can  construct a family of vector potentials for which the corresponding Helmholtz equation  is nonregular $R$-separable.   We give examples in Section~\ref{examples}.

\section[Examples of  nonregular  separability for  Hamilton-Jacobi and Helmholtz equations]{Examples of  nonregular  separability\\ for  Hamilton--Jacobi and Helmholtz equations}\label{examples}

\subsection{Examples of restricted regular separation}

The simplest examples of Theorems~\ref{theorem2} and~\ref{theorem3}, and their extensions to the Helmholtz equation, are those for which regular orthogonal separation already occurs. Thus the ``generalized'' St\"ackel matrix~(\ref{genstackel1}) is a true St\"ackel matrix. Consider f\/irst the separable  Hamilton--Jacobi equation in the form
\[
{\cal H}\equiv\sum_{i=1}^N H^{-2}_i u^2_i+V({\mathbf x})=E.
\]
Here, the metric in the orthogonal separable coordinates $x^i$ is
\[
ds^2=\sum_{i=1}^NH_i^2\big(dx^i\big)^2.
\]
There is an associated St\"ackel matrix and constants of the motion  ${\cal L^{\ell} }$, $2\le \ell\le N$, where ${\cal H}= {\cal L}^{1} $. We apply the side condition ${{\cal L}^N}'\equiv {\cal L}^N-\lambda_N=0$ where $\lambda_N$ is a constant scalar. Note that ${{\cal L}^N}'$ is also a constant of the motion. Now the separation equations become
\begin{gather*}
u_i^2 + \left(v_i\big(x^i\big) +\lambda_N s_{iN}\big(x^i\big)\right)+\sum_{j=1}^{N-1} s_{ij}\big(x^i\big)\lambda_j=0,\qquad
i=1,\dots,N,\qquad \lambda_1=-E.
\end{gather*}
 Our ``restricted constants of the motion''
become
\begin{gather*}
{{\cal L}^{\ell}}'\equiv\sum_{j=1}^N T^{\ell j} \big(u_j^2+v_j+\lambda_N s_{jN}\big),
\end{gather*}
so we have modif\/ied the potential. The side conditions are satisf\/ied automatically. Ef\/fectively, we have restricted our Hamiltonian system to  the $N-1$ dimensional hypersurface ${\cal L}^N=\lambda_N$ and seen that the result is again a Hamiltonian system. Under some circumstances this has a~simple interpretation as a separable system on a manifold of one less dimension.

\begin{example} Consider the Kepler Hamiltonian
\begin{gather*}
{\cal H}={\cal L}^1=p_x^2+p_y^2+p_z^2+\frac{\alpha}{r},\qquad r=\sqrt{x^2+y^2+z^2}
\end{gather*}
in real Euclidean space. The system is separable in spherical coordinates $r$, $\theta$, $\phi$ where
\begin{gather*}
 x=r\sin\theta\cos\phi,\qquad y=r\sin\theta\sin\phi,\qquad z=r\cos\theta,\\
 r\ge 0,\qquad 0\le\theta\le\pi,\qquad 0\le \phi<2\pi.
\end{gather*}
Here
\[
{\cal H}=p_r^2+\frac{{\cal L}^2}{r^2}+\frac{\alpha}{r},\qquad {\cal L}^2=p_\theta^2+\frac{{\cal L}^3}{\sin^2\theta},\qquad {\cal L}^3=p_\phi^2.
\]
We choose the side condition ${\cal L}^3=\lambda_3$. Then the reduced Hamiltonian becomes
\[
{\cal H}'=p_r^2+\frac{p_\theta^2}{r^2}+\frac{\lambda_3}{r^2\sin^2\theta}{+\frac{\alpha}{r}}.
\]
In terms of the new variables $X=r\cos\theta,\ Y=r\sin\theta$ this becomes
\[
{\cal H}'=p_X^2+p_Y^2+\frac{\alpha}{\sqrt{X^2+Y^2}}+\frac{\lambda_3}{Y^2},
\]
a regular separable system in the f\/irst quadrant of the Euclidean plane. In the special case $\lambda_3=0$ the singularity on the $X$-axis disappears and we can extend the system to the full punctured plane. Note that this construction is conceptually distinct from simply restricting a trajectory conf\/ined to a plane.
\end{example}

A similar construction works for the Helmholtz equation. We want eigenfunctions of the separated form $\Psi=\prod\limits_{j=1}^N\Psi^{(j)}(x^j)$. We take separation equations in the form
\begin{gather*}
\partial_\ell^2\Psi+f_\ell\big(x^\ell\big)\partial_\ell\Psi+\left(v_\ell\big(x^\ell\big)-s_{\ell N}\big(x^\ell\big)\lambda_N-\sum_{\alpha=1}^{N-1}s_{\ell \alpha}\big(x^\ell\big)\lambda_\alpha\right)\Psi=0,\qquad \ell=1,\dots ,N.
\end{gather*}
Then we have the restricted eigenvalue equations
\begin{gather*}
 {L^\beta}'\Psi\equiv \sum_{\ell=1}^NT^{\beta\ell}\big(\partial_\ell^2+f_\ell \partial_\ell+v_\ell-s_{\ell N}\lambda_N\big)\Psi=\lambda_\beta\Psi,\qquad \beta=1,\dots,N-1,
\end{gather*}
and the side condition
\begin{gather*}
{L^N}'\Psi\equiv  \sum_{\ell=1}^NT^{N\ell}\big(\partial_\ell^2+f_\ell\partial_\ell+v_\ell-s_{\ell N}\lambda_N\big)\Psi=0.
\end{gather*}
We take ${L^1}'=H'$.
Ef\/fectively, we have restricted our separable quantum system to an eigenspace $\{\Psi :{L^N}\Psi=\lambda_N\Psi\}$ of a symmetry operator. The restriction is again a Hamiltonian system. Again, under some circumstances this has a simple interpretation as a separable quantum system on a manifold of one less dimension.
\begin{example}
Consider the hydrogen atom Hamiltonian,  a constant multiple of
 \begin{gather*}
 {H}={ L}^1=\partial_x^2+\partial_y^2+\partial_z^2+\frac{\alpha}{r},\qquad r=\sqrt{x^2+y^2+z^2}
 \end{gather*}
in real Euclidean space. The system is separable in spherical coordinates $r$, $\theta$, $\phi$:
\[
 H=\partial^2_r+\frac{2}{r}\partial_r+\frac{1}{r^2} L^2+\frac{\alpha}{r},\qquad  L^2=\partial^2_\theta+\cot\theta\partial_\theta+\frac{1}{\sin^2\theta} L^3,\qquad { L}^3=\partial_\phi^2.
 \]
We choose the side condition $ L^3\Psi=\lambda_3\Psi$. Then the reduced Hamiltonian becomes
\[
H'=\partial_r^2+\frac{2}{r}\partial_r+\frac{1}{r^2}\partial_\theta^2+\frac{\cot\theta}{r^2}\partial_\theta
+\frac{\lambda_3}{r^2\sin^2\theta}{+\frac{\alpha}{r}}.
\]
Now we make an $R$-transformation $\Psi=R\Theta$ where $R=\frac{1}{r\sqrt{\sin\theta}}$. Then the eigenvalue equation $H'\Psi=E\Psi$ becomes
\[
\left(\partial_r^2+\frac{1}{r^2}\partial_\theta^2+\frac{\lambda_3+\frac14}{r^2\sin^2\theta}+\frac{\frac 1 4}{r^2}+\frac{\alpha}{r}\right)\Theta=E\Theta,\qquad \partial_\phi^2\Theta=\lambda_3\Theta.
\]

Now write $\Theta=\Phi(\phi)\Xi(r,\theta)$.
In terms of the new variables $X=r\cos\theta$, $Y=r\sin\theta$ the energy equation is
\[
\left(\partial_X^2+\partial_Y^2+\frac{\alpha}{\sqrt{X^2+Y^2}}{+\frac{\frac 1 4}{X^2+Y^2}}+\frac{\lambda_3+\frac14}{Y^2}\right)\Xi=E\Xi,
\]
a regular separable system in the upper half Euclidean plane. If $\lambda_3=-1/4$ this equation can be extended to the punctured plane and
 regarded as a modif\/ication of the hydrogen atom in the plane. Note, however, that due to the $R$ factor, the usual~$L_2$ normalization for eigenstates in Euclidean space  doesn't restrict to the usual $L_2$ normalization for eigenstates in the plane. Thus, in analogy with the virial theorem for the mapping between the Coulomb and pseudo-Coulomb problems~\cite{Thirring}, one needs to check that the bound state spectra are preserved under the restriction.
\end{example}

\subsection{Nonregular separation in 2D and a ``no go'' theorem}

Now we take up the issue of true nonregular separation in a coordinate system in which regular separation is impossible. Let us f\/irst look at 2D examples, a very special case. By using the facts that we can always replace a separable coordinate by an invertible  function of itself, and we can perform  linear transformations on the separation constants without changing the system, we can always assume that the generalized St\"ackel matrix looks like
\[
{\tilde S}= \begin{pmatrix} 1&A(u,v)\\ 1&B(u,v)\end{pmatrix}.
\]
We must require that $A-B\ne 0$ so that the matrix is invertible and that $AB\ne 0$ so that the metric is nondegenerate. Further, by making use of the equivalence relation (\ref{genstackelequiv}) we can put the matrix in canonical form
\begin{gather*}
S= \begin{pmatrix} 1&1\\ 1&f(u,v)\end{pmatrix},
\qquad \text{with inverse} \quad
T=\frac{1}{f-1} \begin{pmatrix} f&-1 \\ -1&1\end{pmatrix},
\end{gather*}
where $f=B/A$. This form will lead to true nonregular separation unless $f$ can be factored as $f(u,v)=U(u)V(v)$, because in that case we can take $B=V(v),\ A=1/U(u)$ and f\/ind an equivalent true St\"ackel matrix.  It is easy to see that this form leads to true nonregular separation for the Hamilton--Jacobi equation. However, in the 2D case the construction fails for Helmholtz equations.

\begin{theorem} For a $2D$ manifold the Helmholtz equation $\Delta \Psi=\lambda \Psi $ never admits true nonregular separation.
\end{theorem}

\begin{proof} In the 2D case, the  Laplacian takes the form
\[
\Delta=\frac{f}{f-1}\left(\partial_{uu}+\frac12\frac{f_u}{f}\partial_u\right)-\frac{1}{f-1}\left(\partial_{vv}-\frac12\frac{f_v}{f}\partial_v\right).
\]
For  nonregular separation it is necessary that $f_u/f$ is a function of $u$ alone and $f_v/f$ is a function of $v$ alone. Thus $f=U(u)V(v)$ and the system admits regular separation. Similarly if $\Psi=e^R\Theta$ for some f\/ixed~$R$, then $R$-separation  implies that there exist functions $U(u)$, $V(v)$ such that
\[
R_u=-\frac14\frac{f_u}{f}+ U(u),\qquad R_v=\frac14\frac{f_v}{f}+V(v).
\]
Since $\partial_v R_u=\partial_u R_v$ it follows easily that $\partial_{uv}\ln f=0$. Thus $f={\hat U}(u){\hat V}(v) $ and
the system must admit regular $R$-separation.
\end{proof}

However, true nonregular separation  may occur when vector potentials are included, even in the 2D case.
The self-adjoint Schr\"odinger equation for a non-relativistic charged particle inf\/luenced by a classical electromagnetic 4-potential, still does not
 admit true nonregular separation. Nonetheless true nonregular separation does occur when the vector potential gives rise to skew-adjoint f\/irst-order terms,
as in a dif\/fusion-convection equation for transport of solute.

  \subsection[Quantum  particle in ${\mathbb R}^2$ with a magnetic field]{Quantum  particle in $\boldsymbol{{\mathbb R}^2}$ with a magnetic f\/ield} \label{vector potential}

 In the Schr\"odinger picture with Cartesian coordinates, the eigenvalue equation for the quantized Hamiltonian of a non-relativistic charged particle interacting with a classical electromagnetic f\/ield, is (e.g.~\cite{BW2004})
\[
 H\Psi=-\frac{(\hbar )^2}{2}\nabla^2\Psi-\frac{i\hbar }{2}[{\bf A.\nabla}+{\bf \nabla . A}({\bf x})]\Psi+\varPhi({\bf x})\Psi=\lambda \Psi.
 \]
 $(A_1,A_2,0)$ is the vector potential for the magnetic f\/ield and $\varPhi$ is the scalar potential for the electric f\/ield.
 We consider the equivalent eigenvalue problem
 \begin{gather}
  H\Psi = \partial_1^2\Psi+\partial_2^2\Psi
  + \frac{i}{2}\big[a^1({\bf x})\partial_1+a^2({\bf x})\partial_2+{\bf \nabla . a}({\bf x})\big]\Psi+\Phi({\bf x})\Psi=E \Psi.
 \label{magSchrod}
 \end{gather}
 We extend this equation to a general curvilinear orthogonal coordinate system. As before,
\[
\Delta=\left[ H_1^{-2}\partial_1^2+H_2^{-2}\partial_2^2+H_1^{-2}\partial_1 \log\frac{H_2}{H_1} \partial_1-H_2^{-2}\partial_2 \log \frac{H_2}{H_1}\partial_2\right].
\]
 In a general orthogonal coordinate system, the scalar divergence, involving Christof\/fel symbol~$\Gamma$, of vector~$\bf a$,  is
 \begin{gather*}
 \operatorname{div} {\bf a} = \frac{\partial  a ^m}{\partial  x ^m} +  a ^m \Gamma ^i_{im}
 =\frac{1}{\sqrt{ g}}\frac{\partial}{\partial x^i} \big( a^i \sqrt{ g}\big)
 \nonumber\\
 \hphantom{\operatorname{div} {\bf a}}{}
 = \partial _i a^i +a^1[\partial_1 \log H_1+\partial_1 \log H_2]+a^2[\partial_2 \log H_1+\partial_2 \log H_2].
\end{gather*}
 The eigenvalue equation for the Hamiltonian is
 \begin{gather*}
 \nonumber
 H\Psi = H_1^{-2}\partial_1^2\Psi+H_2^{-2}\partial_2^2\Psi+\left[H_1^{-2}\partial_1 \log\frac{H_2}{H_1} \right]\partial_1\Psi-
 \left[H_2^{-2}\partial_2 \log \frac{H_2}{H_1}\right]\partial_2\Psi\\
 \hphantom{H\Psi=}{}
  + \frac{i}{2}\big[a^1({\bf x})\partial_1+\partial_1a^1({\bf x})+a^2({\bf x})\partial_2+\partial_2a^2({\bf x})\big]\Psi\\
 \hphantom{H\Psi=}{}
 +\frac{i}{2}\big[a^1\partial_1 \log( H_1)+a^1\partial_1 \log (H_2)+a^2\partial_2 \log(H_1)+a^2\partial_2 \log(H_2)\big] \Psi
  + \Phi({\bf x})\Psi=E \Psi.\nonumber
 \end{gather*}
 Consider $R$-separation
 \[
 \Theta=\Psi e^{-R({\bf x})}=\Theta^{(1)}\big(x^1\big)\Theta^{(2)}\big(x^2\big).
 \]
 Then
 \begin{gather*}
 \nonumber
 \left[H_1^{-2}\partial_1^2+H_2^{-2}\partial_2^2+H_1^{-2}\left\{\partial_1 \log\frac{H_2}{H_1}+2\partial_1R\right\}\partial_1
 +H_2^{-2}\left\{-\partial_2 \log\frac{H_2}{H_1}+2\partial_2R\right\}\partial_2\right] \Theta\\
 \nonumber
\qquad{} +H_1^{-2}\left[\partial_1^2R+(\partial_1R)^2+(\partial_1R)\partial_1\log\frac{H_2}{H_1}\right]\Theta\\
 \nonumber
\qquad{} +H_2^{-2}\left[\partial_2^2R+(\partial_2R)^2-(\partial_2R)\partial_1\log\frac{H_2}{H_1}\right]\Theta+\Phi({\bf x})\Theta\\
 \nonumber
\qquad{} +\frac{i}{2}\big[a^1\partial_1 \Theta+a^2\partial_2 \Theta\big]+\frac{i}{2}\big[a^1\partial_1R+a^2\partial_2R+\partial_1a^1+\partial_2a^2\big]\Theta\\
\qquad{}+ \frac{i}{2}a^1[\partial_1 \log H_1+\partial_1 \log H_2]\Theta+\frac{i}{2}a^2[\partial_2 \log H_1+\partial_2 \log H_2]\Theta=E\Theta.
 \end{gather*}
 Assume the separation equations
 \begin{gather*}
 \partial_1^2\Theta+[f_1(x^1)+ig_1(x^1)]\partial_1\Theta+\left[[v_1(x^1)+iw_1(x^1)]-S_{11}(x^1)E\right]\Theta = 0,\nonumber\\
  \partial_2^2\Theta+[f_2(x^2)+ig_2(x^2)]\partial_2\Theta+\left[[v_2(x^2)+iw_2(x^2)]-S_{21}(x^2)E\right]\Theta = 0, \quad \Rightarrow \nonumber\\
   T_{11}\partial_1^2\Theta+T_{12}\partial_2^2\Theta+T_{11}[f_1+ig_1]\partial_1\Theta+T_{12}[f_2+ig_2]\partial_2\Theta\nonumber \\
 \qquad{} +\left[ T_{11}[v_1+iw_1]+T_{12}[v_2+iw_2] \right]\Theta = E\Theta,\nonumber\\
 T_{21}\partial_1^2\Theta+T_{22}\partial_2^2\Theta+T_{21}[f_1+ig_1]\partial_1\Theta+T_{22}[f_2+ig_2]\partial_2\Theta\nonumber\\
\qquad{}  +\left[T_{21}[v_1+iw_1]+T_{22}[v_2+iw_2]\right]\Theta = 0,   
  \end{gather*}
  where matrix $T$ is the inverse of generalized St\"ackel matrix~$S$. In order for the above eigenvalue equation to be identif\/ied with the Schr\"odinger equation,
 we again require $T_{11}=H_1^{-2}$ and $T_{12}=H_2^{-2}$, and in addition,
    \begin{gather}
  \label{f1eq}
  f_1\big(x^1\big) = \partial_1 \log\frac{H_2}{H_1}+2\partial_1R,\qquad
   f_2\big(x^2\big) = -\partial_2 \log\frac{H_2}{H_1}+2\partial_2R, \\
  \nonumber
  T_{11}v_1 + T_{12}v_2=\Phi+H_1^{-2}\partial_1^2R+H_2^{-2}\partial_2^2R+H_1^{-2}(\partial_1R)^2+H_2^{-2}(\partial_2R)^2\\
\hphantom{T_{11}v_1 + T_{12}v_2=}{}
+ H_1^{-2}(\partial_1R)\partial_1 \log\frac{H_2}{H_1}-H_2^{-2}(\partial_2R)\partial_2 \log\frac{H_2}{H_1},
 \nonumber \\
  g_1\big(x^1\big) = \frac{1}{2}H_1^2a^1,\qquad
   g_2\big(x^2\big) = \frac{1}{2}H_2^2a^2,  \nonumber\\
  \nonumber
  T_{11}w_1 + T_{12}w_2=\frac{1}{2}\big[a^1\partial_1R+a^2\partial_2R+\partial_1a^1+\partial_2a^2\big]\\
\hphantom{T_{11}w_1 + T_{12}w_2=}{}
+ \frac{1}{2}a^1[\partial_1\log H_1+\partial_1 \log H_2]+\frac{1}{2}a^2[\partial_2 \log H_1+\partial_2 \log H_2].
 \label{weq}
  \end{gather}
  From (\ref{f1eq}) 
   we can easily deduce that $\partial_1\partial_2R=0$ and $\partial_1\partial_2 \log(H_2/H_1)=0$.  Hence
 \begin{gather}
  \label{sepcond}
 H_2/H_1 = \Pi_1\big(x^1\big)\Pi_2\big(x^2\big)\qquad  \hbox{and} \qquad R= R_1\big(x^1\big)+R_2\big(x^2\big)
 \end{gather}
 for some functions $\Pi_1$, $\Pi_2$, $R_1$, $R_2$.
 Now assuming the canonical form with $H_1=1$, we have
 \begin{gather*}
 a^1=2g_1\big(x^1\big) \qquad \hbox{and}\qquad
 a^2=2g_2\big(x^2\big)\Pi_2\big(x^2\big)^{-2}\Pi_1\big(x^1\big)^{-2},
 \end{gather*}
whence it follows by separation of variables in (\ref{weq}),  that
\begin{gather*}
 \Pi_1^2[w_1-g_1\partial_1R_1-\partial_1g_1-g_1\partial_1\log \Pi_1] = \gamma_4,\\
 \Pi_2^{-2}[w_2-g_2\partial_2R_2-\partial_2g_2+g_2\partial_2\log \Pi_2] = -\gamma_4,
\end{gather*}
where $\gamma_4$ is constant. By integration,
\begin{gather*}
a^1 = 2\Pi_1^{-1}e^{-R_1}\left[\gamma_5+\int e^{R_1}\big[w_1\Pi_1-\gamma_4\Pi_1^{-1}\big]dx^1\right],\\
 a^2 = 2\Pi_1^{-2}\Pi_2^{-1}e^{-R_2}\left[ \gamma_6+\int e^{R_2}\big[w_2\Pi_2^{-1}+\gamma_4\Pi_2\big]dx^2\right],
\end{gather*}
with $\gamma_4$, $\gamma_5$ and $\gamma_6$ constant.
As a standard test, we f\/irst recover the known possibility of separation in polar coordinates in $E^2$, with
\[
x^1=r,\qquad x^2=\theta, \qquad \Pi_1=r, \qquad \Pi_2=1.
\]
In  (\ref{f1eq}) 
we may choose $f_1=f_2=0$ without any loss of generality to the class of allowable scalar and vector potentials.  Consequently, from (\ref{f1eq}),
$R_1=-\frac{1}{2}\log r$ and $R_2=0$. This yields the potential f\/ields that admit separation of the Schr\"odinger equation
\begin{gather*}
a^1 = \alpha_1(r)=\gamma_5 r^{-1/2}+2\gamma_4r^{-1}+r^{-1/2}\int r^{1/2}w_1(r)dr ,\\
a^2 = \alpha_2(\theta)r^{-2}=2\gamma_6r^{-2} +2r^{-2}\int\gamma_4+w_2(\theta)d\theta,\qquad
\Phi = v_1(r)+r^{-2}v_2(\theta) -\frac 1 4 r^{-2}.
\end{gather*}
The above form for the scalar potential agrees with that known to allow separation of polar coordinates in the Helmholtz operator~\cite{KMJ87}.
When the vector potential~$\bf a$ in~(\ref{magSchrod}) takes values in~${\mathbb R}^2$, the separation condition~(\ref{sepcond}) again leads to the conclusion
 that in this case, generalized St\"ackel matrices do not lead to new separable systems.

\subsection{Solute transport}
The possibilities of new separable equations are broadened if we replace the pure imaginary coef\/f\/icients of f\/irst-order terms in (\ref{magSchrod}) by
f\/irst-order terms with real coef\/f\/icients.  Consider a~solute dif\/fusion-convection equation in Cartesian coordinates,
 \begin{gather*}
 \partial_t\Psi=H\Psi=\partial_1^2\Psi+\partial_2^2\Psi
 -\big[q^1({\bf x})\partial_1+q^2({\bf x})\partial_2\big]\Psi-\mu({\bf x})\Psi .
 \end{gather*}
In this application, $\Psi$ represents the solute concentration,  ${\bf q}({\bf x})$ represents a steady velocity f\/ield of the solvent, and $\mu({\bf x})$ is an adsorption coef\/f\/icient  for removal of solute by the solid substrate of a porous medium or by another (usually solid) component of the mixture. If $\Psi({\bf x})$ satisf\/ies $H\Psi=E\Psi$, then
$e^{Et}\Psi({\bf x})$ is a solution of the time dependent solute equation. In standard applications, $E\le 0$, unless the solid component of the mixture is releasing solute $(\mu({\bf x})<0)$ rather than adsorbing solute.
After making the replacements ${\bf q}=(-i/2){\bf a}$ and $\mu(\mathbf x)=-\Phi(\mathbf x)-(i/2){\bf \nabla. a}$, the solute equation is directly analogous to the Schr\"odinger equation with magnetic f\/ield, except that now $q^j$ must be real, corresponding to~$a^j$  being pure imaginary, $a^j=-iA^j$, $A^j=-2q^j$,
\begin{gather*}
 \partial_t\Psi = H\Psi=\partial_1^2\Psi+\partial_2^2\Psi
 + \frac{1}{2}\big[A^1(\mathbf x)\partial_1+A^2(\mathbf x)\partial_2+\nabla . {\bf A}(\mathbf x)\big]\Psi+\Phi(\mathbf x)\Psi.
 \end{gather*}
 In a general orthogonal coordinate system,
 \begin{gather*}
 \nonumber
 H\Psi = H_1^{-2}\partial_1^2\Psi+H_2^{-2}\partial_2^2\Psi+\left[H_1^{-2}\partial_1 \log\frac{H_2}{H_1} \right]\partial_1\Psi-
 \left[H_2^{-2}\partial_2 \log \frac{H_2}{H_1}\right]\partial_2\Psi\\
\hphantom{H\Psi =}{}
+ \frac{1}{2}\big[A^1({\bf x})\partial_1+\partial_1A^1({\bf x})+A^2({\bf x})\partial_2+\partial_2A^2({\bf x})\big]\Psi+ \frac{1}{2}\big[A^1\partial_1 \log( H_1)\nonumber\\
\hphantom{H\Psi =}{}
+A^1\partial_1 \log (H_2)+A^2\partial_2 \log(H_1)+A^2\partial_2  \log(H_2)\big] \Psi
  +\Phi(\mathbf x)\Psi
  =E \Psi,
 \end{gather*}
consider $R$-separation
 \[
 \Theta=\Psi e^{-R({\bf x})}=\Psi^{(1)} \big(x^1\big)\Psi^{(2)}\big(x^2\big).
 \]
 Then
 \begin{gather*}
 \nonumber
 \left[H_1^{-2}\partial_1^2+H_2^{-2}\partial_2^2+H_1^{-2}\left\{\partial_1 \log \frac{H_2}{H_1}+2\partial_1R\right\}\partial_1
 +H_2^{-2}\left\{-\partial_2  \log \frac{H_2}{H_1}+2\partial_2R\right\}\partial_2\right] \Theta\\
 \nonumber
 \qquad{} + H_1^{-2}\left[\partial_1^2R+(\partial_1R)^2+(\partial_1R)\partial_1 \log \frac{H_2}{H_1}\right]\Theta\\
 \nonumber
 \qquad{} + H_2^{-2}\left[\partial_2^2R+(\partial_2R)^2-(\partial_2R)\partial_1 \log \frac{H_2}{H_1}\right]\Theta+\Phi(\mathbf x)\Theta\\
 \nonumber
 \qquad {} + \frac{1}{2}\big[A^1\partial_1 \Theta+A^2\partial_2 \Theta\big]+\frac{1}{2}\big[A^1\partial_1R+A^2\partial_2R+\partial_1A^1+\partial_2A^2\big]\Theta\\
 \qquad {} + \frac{1}{2}A^1[\partial_1 \log H_1+\partial_1 \log H_2]\Theta+\frac{1}{2}A^2[\partial_2 \log H_1+\partial_2 \log H_2]\Theta=E\Theta.
 \end{gather*}
 Assume the separation equations
 \begin{gather*}
 \partial_1^2\Theta+f_1\big(x^1\big)\partial_1\Theta+\left[v_1\big(x^1\big)-S_{11}\big(x^1\big)E\right]\Theta = 0,\nonumber\\
  \partial_2^2\Theta+f_2\big(x^2\big)\partial_2\Theta+\left[v_2\big(x^2\big)-S_{21}\big(x^2\big)E\right]\Theta = 0,  \qquad \Rightarrow
  \\
  \nonumber
T_{11}\partial_1^2\Theta+T_{12}\partial_2^2\Theta+T_{11}f_1\partial_1\Theta+T_{12}f_2\partial_2\Theta
   + [ T_{11}v_1+T_{12}v_2  ]\Theta = E\Theta,\\
 T_{21}\partial_1^2\Theta+T_{22}\partial_2^2\Theta+T_{21}f_1\partial_1\Theta+T_{22}f_2\partial_2\Theta   + [T_{21}v_1+T_{22}v_2 ]\Theta = 0,
  \end{gather*}
  where matrix $T$ is the inverse of generalized St\"ackel matrix $S$. In order for the above eigenvalue equation to be identif\/ied with the solute equation, we again require $T_{11}=H_1^{-2}$, $T_{12}=H_2^{-2}$. Since the f\/irst-order terms of the target solute equation no longer have imaginary coef\/f\/icients, the identif\/ication of f\/irst-order terms now leads to more general possibilities
   \begin{gather}
  \label{f1Aeq}
 H_1^2q^1= -f_1\big(x^1\big)+\partial_1 \log\frac{H_2}{H_1}+2\partial_1R,\qquad
  H_2^2q^2= -f_2\big(x^2\big)-\partial_2 \log\frac{H_2}{H_1}+2\partial_2R.
   \end{gather}
   By dif\/ferentiating throughout, this implies
   \begin{gather*}
   \partial_1\partial_2\log\frac{H_2}{H_1}+2\partial_1\partial_2 R-\partial_2\big( H_1^2q^1\big)=0, \qquad
   - \partial_1\partial_2\log\frac{H_2}{H_1}+2\partial_1\partial_2 R-\partial_1 \big(H_2^2q^2\big)=0.
   \end{gather*}
    This system is equivalent to
  \begin{gather*}
 4\partial_1\partial_2R-\partial_2\big(H_1^2q^1\big)-\partial_1\big(H_2^2q^2\big)=0,\qquad
 2\partial_1\partial_2\log \frac{H_2}{H_1} -\partial_2\big(H_1^2q^1\big)+\partial_1\big(H_2^2q^2\big)=0.
 \end{gather*}
  Similarly, identif\/ication of the zero-th order terms also leads to a more general condition
  \begin{gather}
  \nonumber
  T_{11}v_1 + T_{12}v_2=-\mu+H_1^{-2}\partial_1^2R+H_2^{-2}\partial_2^2R+H_1^{-2}(\partial_1R)^2+H_2^{-2}(\partial_2R)^2\\
\hphantom{T_{11}v_1 + T_{12}v_2=}{}
+ H_1^{-2}(\partial_1R)\partial_1 \log\frac{H_2}{H_1}-H_2^{-2}(\partial_2R)\partial_2 \log\frac{H_2}{H_1} -q^1\partial_1R-q^2\partial_2R.
   \label{vAform}
  \end{gather}
  In principle, the compatible velocity f\/ield $\bf q$ is recovered from (\ref{f1Aeq})
  after substituting these expressions in (\ref{vAform}) which determines the function~$R$.

  The  condition $H_2/H_1=\Pi_1(x^1)\Pi_2(x^2)$, which is the key condition for the possibility of replacement by a regular St\"ackel matrix construction, is no longer true in general but occurs only if
\[
 \partial_2\big(H_1^2q^1\big)-\partial_1\big(H_2^2q^2\big)=0.
 \]
 In a description with Cartesian coordinates for Euclidean space, this would be exactly the restriction that the solvent velocity f\/ield is irrotational; $\bf \nabla \times \bf q=\bf 0$.
  Consider the canonical form
\[
S^{-1}=T=\frac{1}{f({{\bf x}})-1}\left[
\begin{matrix}
f & -1\\
-1 & 1
\end{matrix}
\right].
\]
Then (\ref{vAform}) may be written as,
 \begin{gather*}
  -fv_1+v_2 = -f\partial_1^2R+\partial_2^2R+f(\partial_1R)^2-(\partial_2R)^2
-f\partial_1R\partial_1\log (-f)+2f\partial_1R\partial_1 \log (1-f)\\
  \nonumber
\hphantom{-fv_1+v_2 =}{}
-f_1\partial_1R +\partial_2 R\partial_2 \log (-f)-2\partial_2R\partial_2 \log (1-f)+f_2\partial_2 R
+(f-1)\mu.
      \end{gather*}
In looking for examples of genuine nonregular separation, we must consider metrics for which $H_2/H_1$ does not separate in variables~$x^1$ and~$x^2$.

\begin{example} We consider an example in 2D Euclidean space. The metric is
\begin{gather*}
 ds^2=dx^2+dy^2=\frac{f(u,v)-1}{f(u,v)}du^2+(1-f(u,v))dv^2,\\
  f(u,v)=-\frac14\left(u+v+\sqrt{(u+v)^2-4}\right)^2.
\end{gather*}
Here
\begin{gather*}
x=(u+v)\cos(\phi-u),\qquad y=(u+v)\sin(\phi-u),\\
\phi= \frac12\left(u+v+\sqrt{(u+v)^2-4}\right)-2\arctan\left(\frac{u+v+\sqrt{(u+v)^2-4}}{2}\right),
\end{gather*}
and $|u+v|\ge 2$.
The Helmholtz equation with vector and scalar potential takes the form
\begin{gather*}
\left(\Delta_2-\frac{f}{f-1}\left(\frac12\frac{f_u}{f}\partial_u-U(u)\right)
+\frac{1}{f-1}\left(\frac{{-}1}{2}\frac{f_v}{f}\partial_v-V(v)\right)\right)\Psi=E\Psi.
\end{gather*}
The separation equations are{\samepage
\[
 (\partial_{uu}+U(u)-E)\Psi^{(1)}(u)=0,\qquad (\partial_{vv}+V(v)-E)\Psi^{(2)}(v)=0,
 \]
with $\Psi=\Psi^{(1)}(u)\Psi^{(2)}(v)$.}

 Now $r^2=x^2+y^2=(u+v)^2$, from which the polar coordinate may be taken to be $r=u+v$. Then
\[
x=r\cos(\phi-u),\qquad y=r\sin(\phi-u),
\]
implying that the polar angle coordinate is
\[
\phi-u=\theta \quad \hbox{(mod 2}\pi).
\]
Also
\[
\phi= \frac12\left(r+\sqrt{r^2-4}\right)-2\arctan\left(\frac{r+\sqrt{r^2-4}}{2}\right).
\]
Note that $\pm\theta$ is an additive component of each of the variables $u=\phi(r)-\theta$ and $v=r-\phi(r)+\theta$. Therefore physically relevant solutions must be periodic with period~$2\pi$ in each of the variables~$u$ and~$v$.
Taking the simplest case $v^1(=U)=0$ and $f^1(=V)=0$, direct  separation is possible with  $R=0$, and from (\ref{f1Aeq}) 
\begin{gather*}
(H_1)^2q^1=-(H_2)^2q^2=1/\sqrt{(u+v)^2-4}=\big[r^2-4\big]^{-0.5}.
\end{gather*}
The velocity f\/ield is def\/ined on the exterior to the circle of radius 2 centred at the origin.
The squared magnitude of velocity is
\begin{gather*}
\big(q^1\big)^2{\bf e}_1.{\bf e}_1+\big(q^2\big)^2{\bf e}_2.{\bf e}_2=H_1^2\big(q^1\big)^2+H_2^2\big(q^2\big)^2=\frac{1}{r^2-4}
\end{gather*}
After taking $E=-\omega^2$, separated solutions for solute concentration may be combined in a~Fourier integral
\[
\Psi=\int_0^{\infty}A(\omega)e^{-\omega^2t} \cos(\omega[u+\delta(\omega)])cos(\omega[v+\varepsilon(\omega)])d\omega,
\]
with amplitude function $A(\omega)$ and phase functions $\delta(\omega)$, $\varepsilon(\omega)$.
 \end{example}

\subsection{Nonregular separation in more than 2 dimensions}

In dimensions greater than two, both true nonregular separation and $R$-separation occur even without an added vector potential.
\begin{example}
This   is a 3D Minkowski space example. The metric is
\[
 ds^2=dt^2-dx^2-dy^2=-\frac{4w^2}{(u^2-v^2)^2}\big(du^2+dv^2\big)+dw^2=H_u^2 du^2+H_v^2dv^2+H_w^2 dw^2.
 \]
Here,
\[
 t=\frac{w}{u^2-v^2}\left(\frac14+\big(u^2+v^2\big)^2\right),
 \qquad x=\frac{w}{u^2-v^2}\left(\frac14-\big(u^2+v^2\big)^2\right),\qquad y=\frac{2wuv}{u^2-v^2},
\]
and we require $u$, $v$ real and $w>0$, $|u|>|v|$.

The Helmholtz equation in 3D Minkowski space is
\[
 \Delta_3\Theta=E\Theta,
\]
where
\[
\Delta_3=\frac{\big(u^2-v^2\big)^2}{4w^2}\left(-\partial_{uu}-\partial_{vv}+\frac{4w^2}{\big(u^2-v^2\big)^2}\partial_{ww}
+\frac{8w}{\big(u^2-v^2\big)^2}\partial_w\right).
\]

 We look for $R$-separable solutions
\[
\Theta=e^R \Psi=e^R \Psi^{(1)}(u)\Psi^{(2)}(v)\Psi^{(3)}(w),\qquad e^R=\frac{1}{w}.
\]
By direct calculation we can establish the operator identity
\[
 e^{-R}\Delta_2  e^R -E =H_u^{-2}(\partial_{uu}+\lambda_2)+H_v^{-2}(\partial_{vv}-\lambda_2)+H_w^{-2}\left(\partial_{ww}-\frac{1}{w^2}-E\right).
 \]
Thus the generalized St\"ackel matrix and its inverse are
\[
 S= \begin{pmatrix} 0 & 1& 1+\dfrac{\big(u^2-v^2\big)^2}{4w^2}\vspace{1mm}\\ 0& -1&-\dfrac{\big(u^2-v^2\big)^2}{4w^2}\vspace{1mm}\\ 1& 0& \dfrac{\big(u^2-v^2\big)^2}{4w^2}\end{pmatrix},\qquad
T= \begin{pmatrix}
-\dfrac{\big(u^2-v^2\big)^2}{4w^2} & -\dfrac{\big(u^2-v^2\big)^2}{4w^2} &1\vspace{1mm}\\-\dfrac{\big(u^2-v^2\big)^2}{4w^2}& -1-\dfrac{\big(u^2-v^2\big)^2}{4w^2}&0\vspace{1mm}\\ 1&1&0
\end{pmatrix}.
\]
In terms of the $\Psi$  functions we have
\begin{gather*}
 L^1\Psi=\left(-\frac{\big(u^2-v^2\big)^2}{4w^2}(\partial_{uu}+\partial_{vv})+\left(\partial_{ww}-\frac{1}{w^2}\right)\right)\Psi=E\Psi,\\
 L^2\Psi=\left(-\frac{\big(u^2-v^2\big)^2}{4w^2}(\partial_{uu}+\partial_{vv})-\partial_{vv}\right)\Psi=-\lambda_2\Psi,\qquad
L^3\Psi=\left(\partial_{uu}+\partial_{vv}\right)\Psi=0.
\end{gather*}
The separation equations are
\begin{gather*}
(\partial_{uu}+\lambda_2)\Psi^{(1)}=0,\qquad (\partial_{vv}-\lambda_2)\Psi^{(2)}=0,\qquad  \left(\partial_{ww}-\frac{1}{w^2}-E\right)\Psi^{(3)}=0.
\end{gather*}
In terms of the $\Theta $ functions the f\/inal separated solution of
$
\Delta_3\Theta=E\Theta
$
is
\[
\Theta =\frac{\Psi^{(1)}(u)\Psi^{(2)}(v)\Psi^{(3)}(w)}{w}.
\]
\end{example}

\begin{example}
Consider the Helmholtz equation in 3D Euclidean space
$
\Delta_3\Theta=E\Theta$.
It is well known that this equation is regular separable in exactly eleven coordinate systems. Moreover true regular $R$-separation does not occur for
any constant curvature space~\cite{ERNIE}.  This example, due originally to Sym \cite{IMA06,PSym,SYM, SSym} and presented from our point of view, shows that nonregular $R$-separation with two side conditions occurs in orthogonal coordinates distinct from the usual eleven.
Consider  Dupin-cyclidic coordinates $u$, $v$, $w$ such that
\begin{gather*}
x=\frac{b^2\cos u\cosh v +(c\cosh v -a\cos u)w}{a\cosh v -c\cos u},\qquad y=\frac{b\sin u(a\cosh v-w)}{a\cosh v -c\cos u},\\
z=\frac{b\sinh v(w-c\cos u)}{a\cosh v -c\cos u}.
\end{gather*}
Here $a$, $b$, $c$ are positive parameters, with $b^2=a^2-c^2$, $c<a$, and $u\in [0,2\pi)$. For any real $v$ we require $c\cos u< w< a\cosh v$, so that $x$, $y$, $z$ are real.
Since
\begin{gather*}
 ds^2=dx^2+dy^2+dz^2=\left(\frac{b(a\cosh v-w)}{a\cosh v-c\cos u}\right)^2 du^2+\left(\frac{b(w-c\cos u)}{a\cosh v -c\cos u}\right)^2dv^2+dw^2\\
\hphantom{ds^2}{}
=H_1^2du^2+H_2^2dv^2+H_3^2 dw^2,
\end{gather*}
these coordinates are orthogonal. Thus
\begin{gather*}
\Delta_3=\partial_{xx}+\partial_{yy} +\partial_{zz}\\
\hphantom{\Delta_3}{}
=  \frac{1}{H_1H_2H_3}\left[\partial_u(H_1^{-1}H_2H_3\partial_u)+\partial_v(H_1^{1}H_2^{-1}H_3\partial_v)+\partial_w(H_1H_2H_3^{-1}\partial_w)\right].
\end{gather*}
 We look for $R$-separable solutions
\[
\Theta=e^R \Psi=e^R \Psi^{(1)}(u)\Psi^{(2)}(v)\Psi^{(3)}(w),\qquad e^R=\frac{1}{\sqrt{(w-c\cos u)(a\cosh v-w)}}.
\]
By direct calculation we can establish the operator identity
\[
 e^{-R}\Delta_3  e^R -E =H_1^{-2}\left(\partial_{uu}+\frac14\right)+H_2^{-2}\left(\partial_{vv}-\frac14\right)+H_3^{-2}(\partial_{ww}-E).
\]

Thus the generalized St\"ackel matrix and its inverse can be taken as
\[
 S= \begin{pmatrix} 0 & 1&0\\ 0&0&1\\1&-H_1^{-2}&-H_2^{-2}\end{pmatrix},\qquad
T= \begin{pmatrix} H_1^{-2}&H_2^{-2}& 1\\1&0&0\\0&1&0\end{pmatrix},
\]
In terms of the $\Psi$ functions we have
\begin{gather*}
L^1\Psi=\left(H_1^{-2}\left(\partial_{uu}+\frac14\right)+H_2^{-2}
\left(\partial_{vv}-\frac14\right)+\partial_{ww}\right)\Psi=E\Psi,\\
L^2\Psi=\left(\partial_{uu}+\frac14\right)\Psi=0,\qquad L^3\Psi=\left(\partial_{vv}-\frac14\right)\Psi=0.
\end{gather*}
The separation equations are
\begin{gather*}
\left(\partial_{uu}+\frac14\right)\Psi^{(1)}=0,\qquad
\left(\partial_{vv}-\frac14\right)\Psi^{(2)}=0,\qquad  (\partial_{ww}-E)\Psi^{(3)}=0,\\
\Theta =\frac{\Psi^{(1)}(u)\Psi^{(2)}(v)\Psi^{(3)}(w)}{\sqrt{(w-c\cos u)(a\cosh v-w)}}.
\end{gather*}
Here we have
\[
\big[L^2,L^1\big]=F_{22}L^2+F_{23}L^3,\qquad \big[L^3,L^1\big]=F_{32}L^2+F_{33}L^3,\qquad \big[L^2,L^3\big]=0,
\]
for f\/irst-order dif\/ferential operators $F_{ij}$, so we have $R$-separation with  two side conditions.
\end{example}

\begin{example}
Again we consider the Helmholtz equation in 3D Euclidean space
but now we choose coordinates  $u$, $v$, $w$ such that
\[
 x=\frac{1}{\sqrt{2}}(v+w)\cos u,\qquad y=\frac{1}{\sqrt{2}}(v+w)\sin u,\qquad z=\frac{1}{\sqrt{2}}(v-w).
 \]
 We take a scalar potential
\[
 {\tilde V}(x,y,z)=\frac{U(u)}{(v+w)^2}+V(v)+W(w).
\]
The metric is
\begin{gather*}
ds^2=dx^2+dy^2+dz^2=\frac12\big[(v+w)^2du^2+dv^2+dw^2\big],\\
 H_u^{-2}=\frac{2}{(v+w)^2},\qquad H_v^{-2}=H_w^{-2}=2.
\end{gather*}
Thus
\begin{gather*}
\Delta_3+{\tilde V}=\partial_{xx}+\partial_{yy} +\partial_{zz}+{\tilde V}=
\frac{1}{H_1H_2H_3}\big[\partial_u(H_1^{-1}H_2H_3\partial_u)+\partial_v\big(H_1^{1}H_2^{-1}H_3\partial_v\big)\\
\hphantom{\Delta_3+{\tilde V}=}{}
+\partial_w\big(H_1H_2H_3^{-1}\partial_w\big)\big]
+\frac{U(u)}{(v+w)^2}+V(v)+W(w).
\end{gather*}
 We look for $R$-separable solutions
\[
\Theta=e^R \Psi=e^R \Psi^{(1)}(u)\Psi^{(2)}(v)\Psi^{(3)}(w),\qquad e^R=\big(2\big[x^2+y^2\big]\big)^{-\frac14}.
\]
By direct calculation we can establish the operator identity
\begin{gather*}
 e^{-R}\Delta_3  e^R -E= \frac{2}{(v+w)^2}(\partial_{uu}+U(u)+1)+2\left(\partial_{vv}+\frac{V(v)}{2}-\lambda_2\right)\\
 \hphantom{e^{-R}\Delta_3  e^R -E=}{}
 +2\left(\partial_{ww}+\frac{W(w)}{2}-\frac{E}{2}+\lambda_2\right).
\end{gather*}
Thus the generalized St\"ackel matrix and its inverse can be taken as
\[
 S= \begin{pmatrix} \frac12 & 1&2-\dfrac{1}{(v+w)^2}\\ 0&-1&-2\\0&0&1\end{pmatrix},\qquad
T= \begin{pmatrix} 2&2& \dfrac{2}{(v+w)^2}\\0&-1&-2\\0&0&1\end{pmatrix},
\]
The separation equations are
\begin{gather*}
(\partial_{uu}+U(u)+1)\Psi^{(1)}=0,\qquad \left(\partial_{vv}+\frac{V(v)}{2}-\lambda_2\right)\Psi^{(2)}=0,\\ \left(\partial_{ww}+\frac{W(w)}{2}-\frac{E}{2}+\lambda_2\right)\Psi^{(3)}=0, \qquad
\Theta =\frac{\Psi^{(1)}(u)\Psi^{(2)}(v)\Psi^{(3)}}{\sqrt{v+w}}.
\end{gather*}
\end{example}

\begin{example}
As a generalization of the previous example, the Euclidean space metrics
\[
ds^2=(U_1(u) w+U_2(u)v +U_3(u))^2du^2+dv^2+dw^2
\]
for arbitrary functions $U_j(u)$ all lead to nonregular separation for Hamilton--Jacobi equations, and in the quantum case,  to nonregular $R$-separation for velocity dependent potentials. The  construction of separation equations and of the generalized St\"ackel matrix
is standard. As an  instance, if
\[
x=F_1(u)v+F_2(u)w,\qquad y=F_3(u)v+F_4(u)w,\qquad z=F_5(u)v+F_6(u)w,
\]
for
\begin{gather*}
 F_1=\sin^2 u,\qquad F_2=\frac{-\cos^3u}{\sqrt{1+\sin^2 u}},\qquad F_3=\sin u\cos u,\\
 F_4=\frac{\sin u\big(1+\cos^2 u\big)}{\sqrt{1+\sin^2 u}}, \qquad F_5=\cos u,\qquad  F_6= \frac{-\sin^2 u}{\sqrt{1+\sin^2 u}},
\end{gather*}
we have
\[
ds^2=dx^2+dy^2+dz^2=\left(\sqrt{1+\sin^2 u} v+\frac{\cos u\big(2+\sin^2 u\big)}{1+\sin^2 u}w\right)^2du^2+dv^2+dw^2.
\]
\end{example}
\begin{example}
The negative constant curvature  metric
\[
ds^2=\frac{dx^2+dy^2+dz^2}{z^2}=\frac{(u+v)^2du^2+dv^2+dw^2}{w^2}
\]
for
\[
 x=(u+v)\sin u+\cos u,\qquad y=-(u+v)\cos u+\sin u,\qquad z=w,
 \]
 leads to nonregular separation for Hamilton--Jacobi equations, and in the quantum case,  to nonregular $R$-separation for velocity dependent potentials~\cite{Cannon}.
\end{example}

\section{Conclusions}\label{conclusions}

We have demonstrated that the
characterization of symmetry related solutions modulo a side condition  for Hamilton--Jacobi and Helmholtz or Schr\"odinger equations coincides
with maximal nonregular separation of variables for which there is a generalized St\"ackel matrix with one arbitrary column. We have also shown that
these systems can be characterized
geometrically, i.e., in a coordinate-free manner. We have demonstrated that there is a structure theory for this type of separation;
it is not just a collection of examples.
 This allows us to obtain new separable solutions for these equations that cannot be obtained by the standard (regular)
methods of separation of variables.

Our work leads to additional questions and possible extensions.
\begin{itemize}\itemsep=0pt
\item Eisenhart showed that the Robertson condition for Helmholtz separabilty of a Hamilton--Jacobi regular separable coordinate system was  the vanishing of the  of\/f-diagonal elements of the Ricci tensor in these coordinates~\cite{EIS48, EIS34}. Is there a corresponding geometrical interpretation of the obstruction problem for nonregular separation?
\item Is there a physical interpretation of the side conditions?
 \item For regular separation of Hamilton--Jacobi and Helmholtz equations all separable systems on $N$-dimensional Euclidean spaces and $N$-spheres are known~\cite{ERNIE}.
Can a similar classif\/ication of separation with a side condition be carried out?
\item Much remains to be done in f\/inding explicit, physically interesting, new solutions via this method.
\item It is clear that separation with a side condition is also appropriate for heat, Laplace and wave equations, where the new solutions are likely to
be more interesting physically. This should be done and a structure theory worked out.
\end{itemize}

\subsection*{Acknowledgements}
This work was partially supported by a grant from the Simons Foundation (\#208754 to Willard Miller, Jr.) and by the Australian Research Council
(grant DP1095044 to G.E.~Prince and P.~Broadbridge).


\pdfbookmark[1]{References}{ref}
\LastPageEnding

\end{document}